\documentclass[pra,preprint,showpacs]{revtex4}
\usepackage[centertags]{amsmath}
\usepackage[koi8-r]{inputenc}
\usepackage{amsfonts}
\usepackage{amssymb}
\usepackage{amsthm} 
\usepackage{newlfont}
\usepackage{epsfig}
\usepackage{amscd}
\usepackage{subcaption}
\usepackage{graphicx}
\usepackage{footnote}
\usepackage{lipsum}
\usepackage{color}
\usepackage{xcolor}
\usepackage{multirow}
\usepackage{amscd}

\newcommand{\beq}{\begin{equation}}
\newcommand{\eeq}{\end{equation}}
\newcommand{\ba}{\begin{array}}
\newcommand{\ea}{\end{array}}
\newcommand{\bea}{\begin{eqnarray}}
\newcommand{\eea}{\end{eqnarray}}
\begin{document}

\begin{center}
{\large \sc \bf {$M$-neighbor approximation in one-qubit state transfer along zigzag { and alternating}  spin-1/2 chains.
}}

\vskip 15pt

E.B.Fel'dman and A.I.Zenchuk

\vskip 8pt

Institute of Problems of Chemical Physics, RAS,
Chernogolovka, Moscow reg., 142432, Russia.

\end{center}

\begin{abstract}
We consider the $M$-neighbor approximation in the problem of one-qubit pure state transfer along the $N$-node  zigzag and alternating
spin chains governed by the $XXZ$-Hamiltonian with the dipole-dipole interaction. We show that always $M>1$, i.e., the nearest neighbor approximation is not applicable to such interaction. Moreover, only all-node interaction  ($M=N-1$) 
properly describes the dynamics in the alternating chain.  We reveal the region in the parameter space characterizing the chain geometry and orientation which provide the high-probability state-transfer. The optimal state-transfer probability and appropriate time instant for the zigzag and  alternating chains are compared.
\end{abstract}


\maketitle

\section{Introduction}
\label{Section:introduction}

The problem of state transfer via the spin-1/2 chain  is a popular problem first formulated by Bose \cite{Bose}. Although photons become a most {admitted} carriers of information  in long distance communication \cite{PBGWK,PBGWK2,DLMRKBPVZBW}, the short distance state transfer can be based on other carriers, among which the  spin-excitation  is acknowledged \cite{PSB,LH}. Such short communication lines are applicable to transfer states between different blocks  of a quantum device. Set of optimizations of a spin transfer line were proposed. Especially we pick out lines 
supporting the perfect state transfer \cite{CDEL,KS} which can be achieved in the $XX$-chain with nearest neighbor interaction. { 
However, the perfect state transfer is not reliable in practical realization because it can be easily destroyed by the Hamiltonian perturbation as  shown in Refs.\cite{CRMF,ZASO,ZASO2,ZASO3}.
It was demonstrated in those references that  the high-probability state transfer  \cite{GKMT},
which, in particular, is prompted by the space-symmetry of the chain \cite{KS},
is more practical being robust with respect to the Hamiltonian perturbations. 
 Therefore it can be considered as an alternative to the perfect state transfer. 
 The high-probability state transfer can be  based on two 
different  approaches. The first one is the state-transfer control through  the coupling constants between the end-nodes and body of a quantum chain (the weak end-bond model) \cite{WLKGGB,GKMT,ZASO2}. The second approach uses the local-field control \cite{DZ_2010}. Both approaches were implemented in the  quantum router in Ref.\cite{PLAPG}, and in multi-user quantum communication line \cite{FZ_2009,YB}.
The weak end-node model was also used in the probabilistic measurement-based state transfer along the noisy chain \cite{BO}. In addition, the model with two pairs of symmetric controlling boundary bonds was  investigated  as well \cite{ABCVV}. Another attractive model used in developing  state transfer protocols is the 
 alternating spin chain, Refs.\cite{FR_2005,KF_2006,VGIZ}. We shall also mention  $d$-level state transfer protocols  \cite{BK,JSTB} and studies of state propagation in $2D$ spin  lattice \cite{LPRA} and in quantum networks \cite{LMSSLLR}.
Lately, state transfer protocols were  generalized to perform the multi-qubit state transfer \cite{HL,YB2} and creation \cite{FPZ_2021},  and multi-excitation state transfer via the perturbative method \cite{CSLA}. Finally, we mention the spin transistor as a quantum device which can control the information flow between the sender and receiver \cite{YBB2}.

{ 
 As a characteristic of   an arbitrary one-qubit pure state transfer  
along the spin-chain  the  fidelity $F$  was proposed in Ref.\cite{Bose}.
Averaged over the pure initial states of the first spin (sender)} (provided that  only one-excitation subspace of quantum states is involved in the spin dynamics) } this fidelity  
can be expressed in terms of the probability amplitude $p_{1N}$ of excited state transfer from the 1st to the last ($N$th) spin : 
\begin{eqnarray}
\label{F}
F=\frac{|p_{1N}|^2}{6}+\frac{|p_{1N}|}{3} + \frac{1}{2},\;\; p_{1N}(t) = P(t) e^{i \phi(t)},\;\;0\le P\le 1,\;\;0\le \phi\le 2 \pi.
\end{eqnarray}
 Therefore we consider the excited state-transfer probability $|p_{1N}|\equiv P$ rather then the fidelity as a characteristics of state transfer.

 {
Concluding the brief review on quantum state-transfer protocols, we  shall remark that the above mentioned space symmetry supporting the high-probability state transfer  is required  up to the robustness with respect to the Hamiltonian perturbations. 
In reality the high-probability state transfer is realizable in much wider class of disordered model beyond the symmetrical ones if only the mentioned disordering can be considered as a perturbation of some symmetrical model. This was demonstrated  for the weak end-bond model, for instance, in Refs. \cite{ZASO3,AML}.}

{
Talking about quantum state transfer  we have to base this process on 
entanglement in the system as a fundamental concept of quantum information science \cite{HW,Wootters,Peres,NCh,AFOV,HHHH}.  For instance, the entanglement between the end-nodes  was considered in \cite{VBR,VGIZ}. In Ref.\cite{YBB}, the entanglement was treated as a resource for reaching the almost perfect state transfer. However, the direct end-to-end entanglement is not completely responsible for the quantum information  transfer as was demonstrated in Ref.\cite{DZ_2017}. A possible scenario of entanglement propagation along a spin chain is so-called relay entanglement \cite{DZ_2018} which provides ''non-instantaneous'' end-to-end entanglement. 
}

{
We recall that the nearest-neighbor interaction is a very popular model in describing the spin evolution governed by $XX$, $XY$ and $XXZ$ Hamiltonians \cite{CDEL,KS,FKZ_2016,FBE_1998}.  { It is remarkable that} most papers quoted above are based on the nearest-neighbor interaction \cite{WLKGGB,VBR,VGIZ,YBB,PLAPG,BO,CSLA,YB,YB2,BK}. In particular, that model allows to use the Jordan-Wigner transformation \cite{JW,CG} to describe the spin-evolution in a large quantum system. 
However, the physical nature of  the nearest-neighbor interaction usually remains beyond discussions in most of the papers. 
In particular, the problem of reducing the dipole-dipole interaction  among all particles in a spin system  to interaction  between just nearest spins has  not been deeply studied. 
{ {In other words, there is a lack of papers comparing the spin evolution governed by the nearest-neighbour and all-node  dipole-dipole interaction.   
Nevertheless, sometimes this problems attracts attention  in  literature. For instance, in  Ref.\cite{CRC},  
the nearest-neighbor interaction was successfully applied to approximate { (up to certain degree)} the evolution of  0- and 2-order coherence intensities in MQ NMR experiment over the spin system with dipole-dipole interaction.

On the contrary, it was shown 
in Ref.\cite{FKZ_2010} that the spin dynamics in the process of quantum state transfer along the chain with dipole-dipole interactions governed  by either $XX$ or $XXZ$ Hamiltonian using the nearest-neighbor interactions  significantly differs from the dynamics governed by the above Hamiltonians involving all-node interactions. The most remarkable difference is in the case of dynamics governed by the $XXZ$ Hamiltonian, when the state-transfer time-interval  is several order longer in the case of nearest neighbor interaction for 10-node spin chain. The time-dependence of the probability amplitude is also quite different.}
{The result in Ref.\cite{FKZ_2010} and the result of such comparison discussed below signify that nearest-neighbour dipole-dipole interaction can not serve as an  approximation to all-node dipole-dipole interaction.} Of course, this conclusion does not reduce the significance of studying the nearest-neighbor models.  Been inapplicable to the spin-systems with dipole-dipole interactions, the nearest-neighbor approximation is satisfactory in the case of fast-decaying exchange interaction where the coupling constants  decrease very fast with the distance. Therefore the nearest-neighbor models remain  of great importance.}

All the above prompts us to answer the following question. Whether the approximation of $M$ ($M>1$)  nearest neighbor ($M$-neighbor approximation) can be used to properly describe the evolution of a spin system with dipole-dipole interaction among all nodes? We show that, in certain cases (but not always), the answer to this question is positive. 

We study the problem of applicability of $M$-neighbor approximation to the state-transfer along the zigzag and alternating chains. The interest to the zigzag chain is prompted by the earlier observations that }
the geometry of a spin system can significantly effect on characteristics of state transfer 
(fidelity and appropriate time instant). For instance, in \cite{DFZ_2009}, the rectangular and parallelepiped  configurations where studied and their advantage in comparison with 1D-chains was demonstrated. Such geometry allows to compactify spin communication lines serving to spread  quantum state among set of receivers. 
The revealed advantage of higher-dimensional configurations 
stimulates our further study of the effect of spin-system geometry on the quantum state transfer. 

We consider  the two-dimensional configuration represented by a zigzag  chain and 
study the characteristics of the state transfer (the probability of the excited state transfer and 
corresponding time-interval) for various values of chain parameters (the direction of the external magnetic field and the geometric parameter responsible for the angle formed by three  neighboring spins) using XXZ-Hamiltonian with all-node dipole-dipole interaction. We reveal the region in the space of these two parameters which provide the large value of state-transfer probability. Then we turn to the $M$-neighbor approximation and reveal the appropriate parameter $M$. Then we perform analogous study for the  state transfer along the alternating chain and compare 
 the characteristics of the state-transfer along the zigzag chain with the appropriate characteristics of the  state-transfer  along the more traditional  alternating chain  of the same horizontal length and reveal  privileges and disadvantages for both of them.

Finally, we notice that the spin chain is not just a theoretical concept.  
As for the candidates for homogeneous spin chain, some nature crystals, such as 
 hydroxy- and fluorapatite crystals \cite{H,VLC,CY}, have chains of $^1H$ and $^{19}F$ suitable for that purpose. 
 Although those chains are not completely isolated from other nuclei in the crystal  and therefore 
they are not perfect one-dimensional spin-1/2 chains.
The zigzag chain is  revealed in the structure of other crystals, for instance, $^1H$ nuclei in hambergite
($Be_2 BO_3 (OH)$) crystal, whose  crystal
structure and crystal chemistry were explored in \cite{Z,ZPM}, and the  experimental investigation of such a crystal via NMR was performed in \cite{BFKLVV}.  

The paper is organized as follows. In Sec.\ref{Section:ev} we describe the XXZ-Hamiltonian with dipole-dipole interaction and  describe the geometry and orientation with respect to the external magnetic field  of the zigzag  
and alternating chains. We also introduce characteristics of the state transfer along those chains.  
In Sec.\ref{Section:Examples} we, first of all, consider the short 4-node chain to compare the evolution of  the state-transfer probability for different number $M$ of nearest interacting  neighbors. Then the state transfer along the long zigzag chain with  both odd and even number of nodes and along the alternating chain with even number of nodes  is analyzed. Conclusions are given in Sec.\ref{Section:conclusions}.
 
\section{End-to-end state transfer with XXZ-Hamiltonian}
\label{Section:ev}

We use the  $XXZ$ Hamiltonian taking into account the dipole-dipole interactions among up to $M$ nearest spins ($M=1$ means the nearest neighbor interaction, $M=N-1$ means the all node interaction):
\begin{eqnarray}
\label{XXZ}
H_{M}=\sum_{i=1}^{N-1}\sum_{j=i+1}^{\min(i+M,N)}
\frac{\gamma^2 \hbar}{2 r^3_{ij}} (3 \cos^2\varphi_{ij}-1)(I_{xi}I_{xj}+I_{yi}I_{yj} - 2I_{zi}I_{zj} ).
\end{eqnarray}
Here $I_{\alpha i}$, $\alpha=x,y,z$, are the operators of the  $\alpha$-projections of the $i$th spin momentum, $\varphi_{ij}$ is the angle between the vector $\vec{r}_{ij}$ and magnetic field $\vec H$, $\gamma$ is the gyromagnetic ratio  and $\hbar$ is the Plank constant. 
As noted in the Introduction, to characterize the fidelity averaged over all initial pure states of the 1-qubit sender, it is enough to consider the transfer of the 1-qubit excited state. 
Therefore, we consider the evolution of the excited state of the  first  spin along the spin-1/2 chain.
Thus, the initial state is 
\begin{eqnarray}\label{in}
|\psi_0\rangle=|1\rangle,
\end{eqnarray}
where $|n\rangle$ means the $n$th excited spin. 
Since the initial state is a one-excitation state, the evolution of this state is described by the
1-excitation block of the evolution operator,
\begin{eqnarray}
V^{(1)}_M(t)=e^{-i H^{(1)}_M t},
\end{eqnarray}
where $M$ is the number of interacting neighbors in Eq.(\ref{XXZ}).
Then the probability of such state transfer from the first to the last node of the chain via $M$-neighbor approximation 
reads
\begin{eqnarray}
p_M(t)=|\langle N|V_M^{(1)}(t)|1\rangle|^2,\;\;  p(t)\equiv p_{N-1}(t).
\end{eqnarray}
Hereafter, unless otherwise specified, instead of $t$, we use the dimensionless time 
\begin{eqnarray}\label{tau}
\tau =\frac{ \gamma^2\hbar}{\Delta^3} t,
\end{eqnarray}
where $2 \Delta$ is the  distance between two nearest odd nodes, see Fig.\ref{Fig:Z}.

\subsection{Zigzag and alternating spin chains}
\label{Section:dm}

\begin{figure*}[!]
\hspace{-2cm}
\begin{subfigure}[h]{0.8\textwidth}
\includegraphics[scale=0.35]{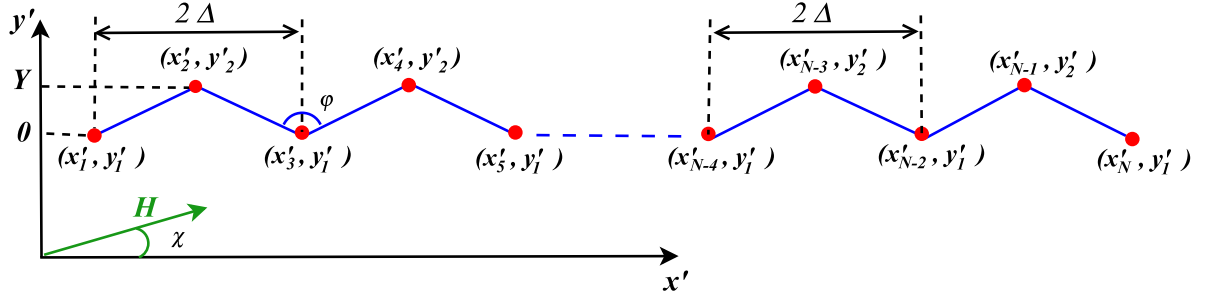}
 \caption{}
\end{subfigure}
\hspace{-2cm}
\begin{subfigure}[h]{0.8\linewidth}
\includegraphics[scale=0.35]{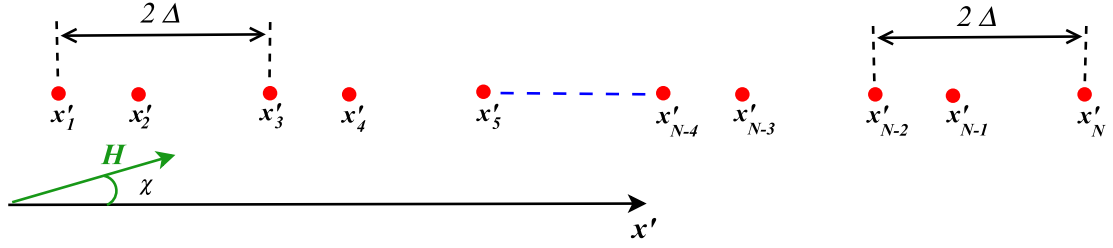}
 \caption{}
\end{subfigure}
\caption{Two schemes used in this paper. (a) Zigzag spin-1/2 chain, $\varphi$ is the angle between directions from a given spin to two nearest nodes. (b) Alternating spin-1/2 chain. $N$ is odd in both cases.  $x_i$ and $y_i$ are the coordinates of the $i$th node. $2 \Delta$ is the distance between two nearest odd nodes, $H$ is the external magnetic field, $\chi$ is the angle between $\vec H$ and $x'$-direction. }
  \label{Fig:Z} 
\end{figure*}

 Here we consider a zigzag, Fig.\ref{Fig:Z}a,  and alternating, Fig.\ref{Fig:Z}b, spin chains
and compare parameters of the one-qubit excited state transfer (probability and appropriate  time instant)  
from the first to the last spin of a chain in both cases. 
We note that the coordinates $x$, $y$ and $z$ in the subscripts of Eq.(\ref{XXZ}) are related with the direction of the magnetic field (which  is $z$-directed). To characterize the positions of spins we introduce the 
system of coordinates $(x',y')$, which is shown in Fig.\ref{Fig:Z}.
Let $j$th spin have coordinates $(x'_j,y'_j)$, the magnetic field ${\mathbf H}$ be directed at the angle $\chi$ to the chain axis.
Then we have 
\begin{eqnarray}\label{coord1}
&&{\mathbf n}=(\cos\chi,\sin\chi),\;\; |{\mathbf n}|=1,\\\label{coord2}
&&
{\mathbf r}_{ij} = (x'_j-x'_i, y'_j-y'_i),\\\label{coord3}
&&
r_{ij} = \sqrt{(x'_j-x'_i)^2 +(y'_j-y'_i)^2 },
\\\label{coord4}
&&
\cos\varphi_{ij}=\frac{(x'_j-x'_i) \cos \chi+(y'_j-y'_i) \sin \chi}{r_{ij}}.
\end{eqnarray}
The $N$-qubit communication line consists of the 1-qubit sender ($S$, the 1st node), $N-2$-qubit transmission line ($TL$) and  1-qubit 
receiver ($R$, the $N$th node).
  
\paragraph{Zigzag spin chain, Fig.\ref{Fig:Z}a.}
We set $x_1'=0$ and
\begin{eqnarray}\label{xy}
x'_j = \Delta (j-1),\;\;y'_j =\left\{\begin{array}{ll}
0,&j=1,3,5,\dots \cr
Y,& j=2,4,6,\dots
\end{array}
\right. .
\end{eqnarray}
We  use $Y$ and $\chi$ as two parameters characterizing, respectively,  the chain geometry and orientation with respect to the strong external magnetic field. 
The length $L_Z$ of the $N$-node  zigzag chain is $L_Z=(N-1)\Delta$ for both 
 odd- and even-node  chains.

\paragraph{Alternating spin chain, Fig.\ref{Fig:Z}b.}

Instead of (\ref{xy}), we have 
\begin{eqnarray}\label{altxy}
x'_j  =\left\{\begin{array}{ll}
 \Delta (j-1),&j=1,3,5,\dots \cr
x'_{j-1}+ \alpha \Delta,& j=2,4,6,\dots
\end{array}
\right. , \;\;0<\alpha<2,
\end{eqnarray}
where $\alpha$ is the alternation parameter, $\alpha=1$ for the homogeneous chain.
Then 
eqs. (\ref{coord2}) - (\ref{coord4}) reduce to 
\begin{eqnarray}\label{altcoord2}
&&
{\mathbf r}_{ij} = (x'_j-x'_i,0),\\\label{altcoord3}
&&
r_{ij} = |x'_j-x'_i|,
\\\label{altcoord4}
&&
\cos\varphi_{ij}= \cos \chi.
\end{eqnarray}
Eq.(\ref{altcoord4}) means that the factor $(3 \cos^2 \varphi_{ij} -1)$ in the coupling constant does not depend on $i$ and $j$, unlike the zigzag chain. 
Therefore, instead of (\ref{tau}),  we can use the dimensionless time $\tau$.
\begin{eqnarray}\label{tau2}
\tau= \frac{ \gamma^2\hbar (3\cos^2 \chi-1)}{\Delta^3} t
\end{eqnarray}
and take into account that the state-transfer probability is invariant with respect to the inversion of $\tau$. Therefore, the sign  of $(3\cos^2 \chi-1)$ does not effect the state-transfer characteristics.
Remark that the threshold value
\begin{eqnarray}
\cos\chi = \frac{1}{\sqrt{3}}
\end{eqnarray} 
destroys the state transfer since all the coupling constants equal zero.
The length $L_A$ of the alternating chain is
\begin{eqnarray}
L_A= \left\{
\begin{array}{ll}
( N-1) \Delta, & {\mbox{odd}} \;\;N\cr
(N-2)\Delta + \alpha\Delta,& {\mbox{even}}\;\;N
\end{array}\right..
\end{eqnarray}
To compare the characteristics of the state transfer along the zigzag and alternating chains we have to consider the chains of equal lengths. For the chains with odd $N$  we have 
\begin{eqnarray}
&&
L_Z=L_A.
\end{eqnarray}
For the chains with even $N$ we have 
\begin{eqnarray}
L_Z-L_A =(1-\alpha)\Delta.
\end{eqnarray}
But if $N\gg 1$, then $L_Z\approx L_A$.

\subsection{Characteristics of state transfer}
Let  $Z$ be the list of parameters, describing the chain geometry and orientation with respect to the external magnetic field:
\begin{eqnarray}\label{Z}
Z=\left\{
\begin{array}{ll}
\{Y, \chi \}  & -\;\; {\mbox{zigzag chain}}\cr
\alpha & -\;\; {\mbox{alternating chain}}
\end{array}
\right. .
\end{eqnarray}
{As the first characteristics, we consider the maximal value $p^{(max)}(T,Z)$ of the probability $p_{N-1}(t,Z)$  of the excited state transfer in the case of all node interaction  and appropriate time 
instant $\tau^{(max)}(T,Z)$ inside of  some time interval  $T$ (which must be fixed conventionally):
\begin{eqnarray}\label{maxp}
p^{(max)}(T,Z) = \max_{0\le \tau \le T} p_{N-1}(\tau,Z) = p_{N-1}(\tau^{(max)}(T,Z),Z).
\end{eqnarray}
}

The second characteristics is needed to reveal  such $M$  in Eq.(\ref{XXZ}) which provides accurate enough approximation to the real spin dynamics. For this aim we introduce 
 the  integral of $p_M(\tau)$  over the time interval $T$, 
\begin{eqnarray}\label{JM}
J_M(T,Z) = \frac{1}{T}\int_0^T p_M(\tau,Z) d\tau.
\end{eqnarray}
Using this integral, we can calculate the ratio 
\begin{eqnarray}\label{ratio}
J_{M,N-1}(T,Z) = \frac{\displaystyle\int_0^T |p_{N-1}(\tau,Z)-p_M(\tau,Z)| d\tau}{J_{N-1}(T,Z)}.
\end{eqnarray}
characterizing the deviation of the approximated value $p_M$ from the correct probability $p\equiv p_{N-1}$. 
Using ratio  (\ref{ratio}) we find  such {minimal value  ${\cal{M}} ={\cal{M}}(T,Z)$  that
\begin{eqnarray}\label{vare1}
&&
J_{M,N-1}(T,Z)\le \varepsilon, \;\;M\ge {\cal{M}},\\\nonumber
&&
J_{M,N-1}(T,Z)> \varepsilon, \;\;M< {\cal{M}},
\end{eqnarray}}
where $\varepsilon$ is desired precision of ${\cal{M}}$-node approximation. Below we set
\begin{eqnarray}\label{vare2}
\varepsilon_1=0.01. 
\end{eqnarray}

\section{Examples}
\label{Section:Examples}
\subsection{Short homogeneous chain}

\begin{figure*}[!]
\epsfig{file=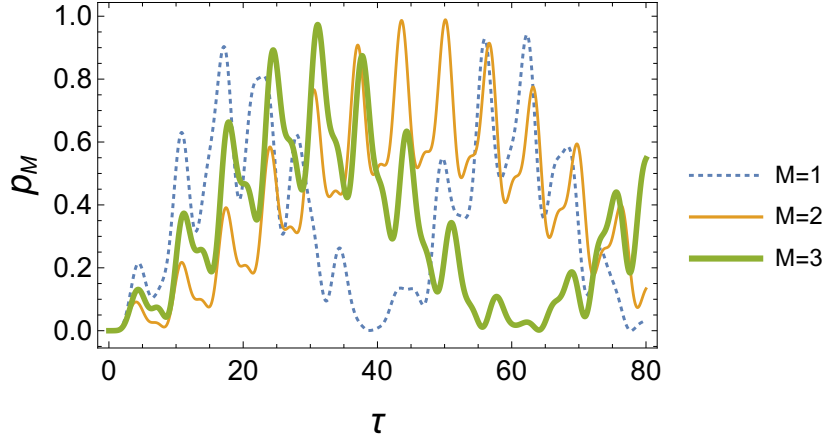,
  scale=0.7
   ,angle=0
} 
\caption{The evolution of  probability $p_M$ of an excited state transfer along the four-qubit chain 
with   $M=1$,  $M=2$ and $M=3$.}
\label{Fig:short}
\end{figure*}

The spin evolution significantly depends on the number $M$ of interacting neighbors in the Hamiltonian (\ref{XXZ}). In fact, let us consider the evolution of  the four-qubit homogeneous chain  (either $Y=0$ in the zigzag chain or $\alpha=1$ in the alternating chain) with $\chi=0$ and different $M$ varying from 1 to 3 in Hamiltonian (\ref{XXZ}). The probability evolution $p_M(\tau)$ is shown in Fig.\ref{Fig:short} for these three cases. We see that all three graphs significantly differ from each other. Therefore, the evolution under the Hamiltonian with $M=1$ (nearest-neighbor approximation) or $M=2$  can not approximate the probability evolution under the Hamiltonian with all-node interaction ($M=3$) and thus is not applicable to study the state-transfer along the spin chain with dipole-dipole interactions.

Thus, including interactions among remote nodes in Hamiltonian  (\ref{XXZ}) is important for obtaining correct result. 
 However, depending on the parameters $Y$ and $\chi$ in the zigzag chain, the minimal number $M>1$ of interacting spins which provides the acceptable approximation to the all-node interaction  may be less then the maximal possible $M=N-1$. 
Therefore, although the approximation of nearest-neighbor interaction does not work, the approximation of $M$-neighbor interaction can be used in certain cases. 

\subsection{Long zigzag chain with even and odd number of spins, Fig.\ref{Fig:Z}a}

To characterize the state transfer, we have to fix the dimensionless    time interval $T$ for state registration. 
We recall that the time instant for  state registration at the receiver of a homogeneous spin chain governed by the $XX$-Hamiltonian 
is usually $\sim N$ (for instance, see  \cite{BZ_2015,FKZ_2016}). However,  simulations of spin dynamics governed by the $XXZ$ Hamiltonian requires longer time of state propagation and $T$ depends on the parity of the zigzag chain. 
Inside of the interval $T$, the probability $p(t)$  can take the bell-shaped form with large amplitude without fast oscillations.  The later  is important for reliability of state registration. 
 
\subsubsection{Odd-node chain: $N=41$.}
\label{Section:zigzag41}

\begin{figure*}[!]
\hspace{-1cm}
\begin{subfigure}[h]{0.5\textwidth}
\includegraphics[scale=0.6]{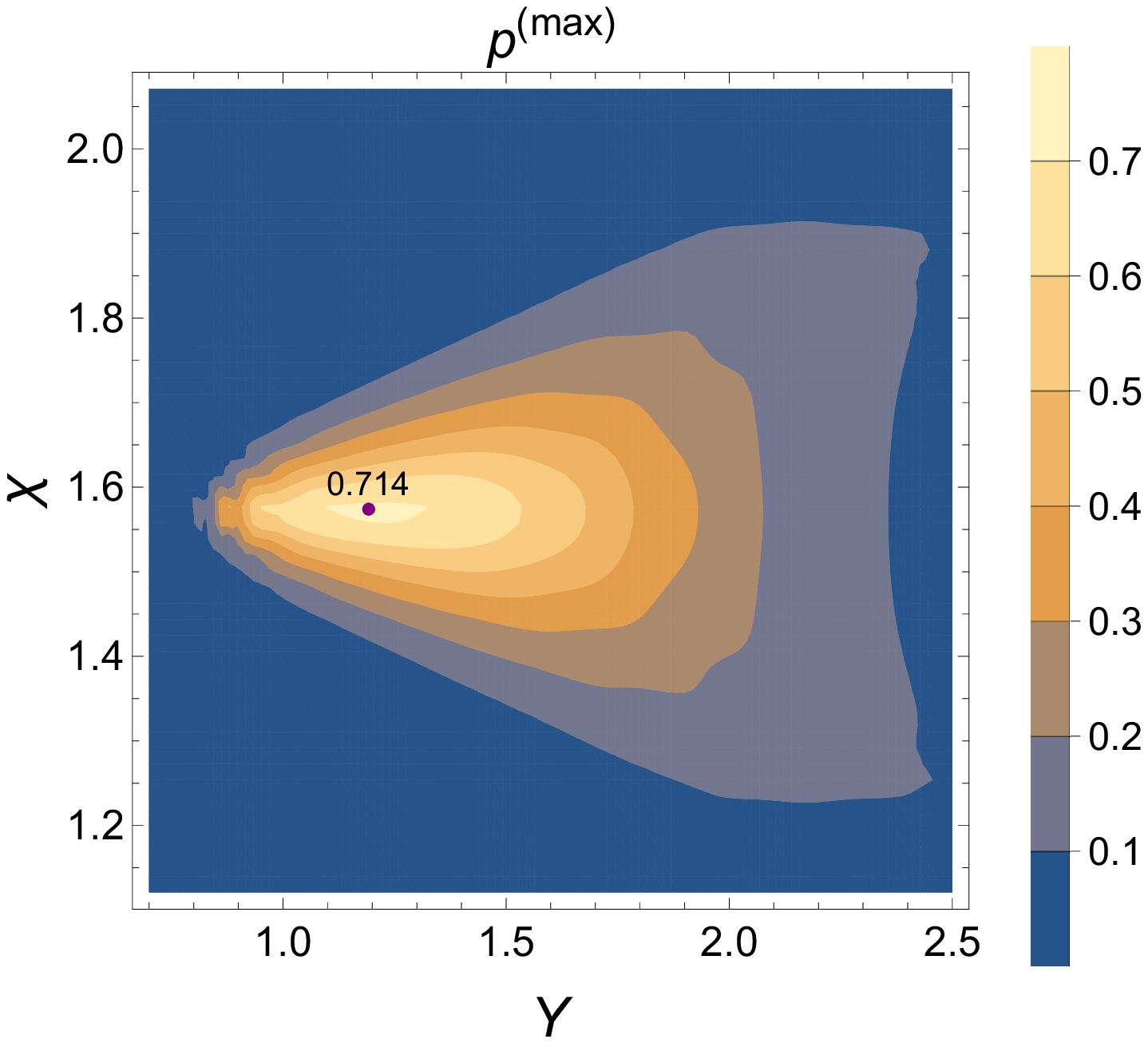}
 \caption{}
\end{subfigure}
\hfill
\begin{subfigure}[h]{0.5\linewidth}
\includegraphics[scale=0.6]{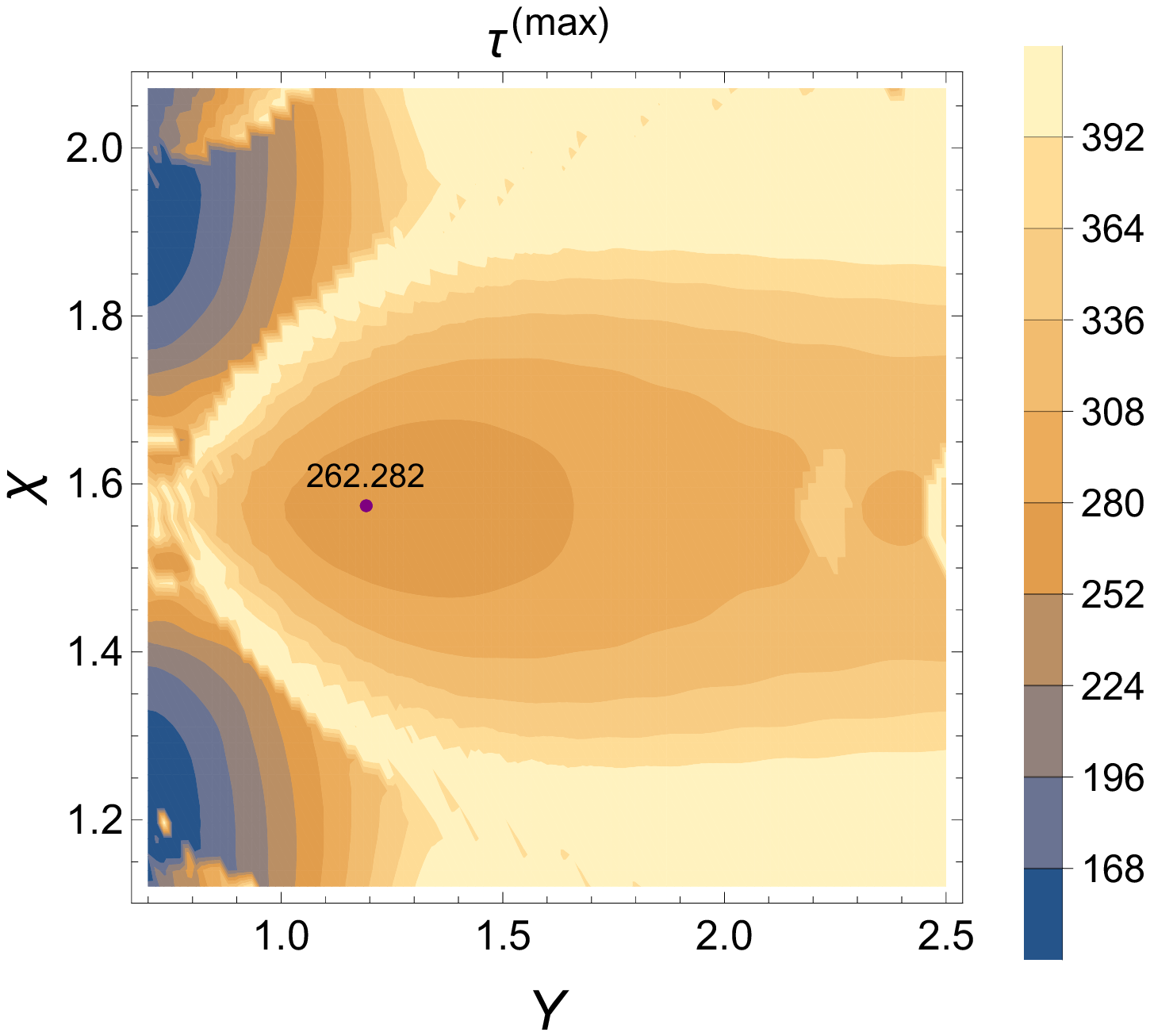}
 \caption{}
\end{subfigure}
\caption{The zigzag chain of $N=41$ nodes. (a) The probability $p^{(max)}$ and (b) the appropriate time instant $\tau^{(max)}$ as  functions 
of $Y$ and $\chi$. The probability $p^{(max)}$ reaches its maximum value { $p^{(opt)}\equiv  p^{(max)} (T, Y^{(opt)}, \chi^{(opt)})=0.714$ at $(Y^{(opt)},\chi^{(opt)})=(1.192, 1.574)$, the appropriate time instant $\tau^{(opt)}\equiv \tau^{(max)}(T, Y^{(opt)}, \chi^{(opt)}) =262.282$.} }
\label{Fig:p41}
\end{figure*}

Numerical simulations show that the probability  $p^{(max)}$ defined in Eq.(\ref{maxp}) takes large values 
($0.1\lesssim p^{(max)} \lesssim 0.7$) inside of the $(Y,\chi)$-region restricted by the conditions
\begin{eqnarray}\label{Ychi41}
0.80\lesssim Y\lesssim 2.45,\;\; 1.23\lesssim \chi \lesssim 1.92
\end{eqnarray}
for   $T=10 N =410$ as shown in  
Fig.\ref{Fig:p41}a.  
The appropriate time instant $\tau^{(max)}$ (see Eq.(\ref{maxp})) is in the interval
\begin{eqnarray}\label{T41}
250 \lesssim \tau^{(max)} \lesssim 400,
\end{eqnarray}
{see Fig.\ref{Fig:p41}b. We call the optimal parameters $Y^{(opt)}$ and $\chi^{(opt)}$ such parameters that maximize $p^{(max)}$ in Fig.\ref{Fig:p41}a. 
The symmetry with respect to $\chi=\frac{\pi}{2}$ in Fig.\ref{Fig:p41} is provided by the mirror-symmetry of the odd-node zigzag chain, i.e., the symmetry with respect to the 
exchange of the $i$th and $(N-i+1)$th nodes  ($i \leftrightarrow N-i+1$).

Now we find ${\cal{M}}$ according to  formulae (\ref{JM}) - (\ref{vare2}). 
The picture of ${\cal{M}}$ over the selected region (\ref{Ychi41}) on the plane $(Y,\chi)$ is shown in Fig.\ref{Fig:MIN41}a, and the 
picture of the integral $J_{N-1}$ over the same region is shown in Fig.\ref{Fig:MIN41}b.  In most cases of large $p^{(max)}$ we have ${\cal{M}}\gtrsim 12$. We found ${\cal{M}}^{(opt)}=12$ at the optimal point 
$(Y^{(opt)}, \chi^{(opt)})$. 
\begin{figure*}[!]
\hspace{-1cm}
\begin{subfigure}[h]{0.5\textwidth}
\includegraphics[scale=0.6]{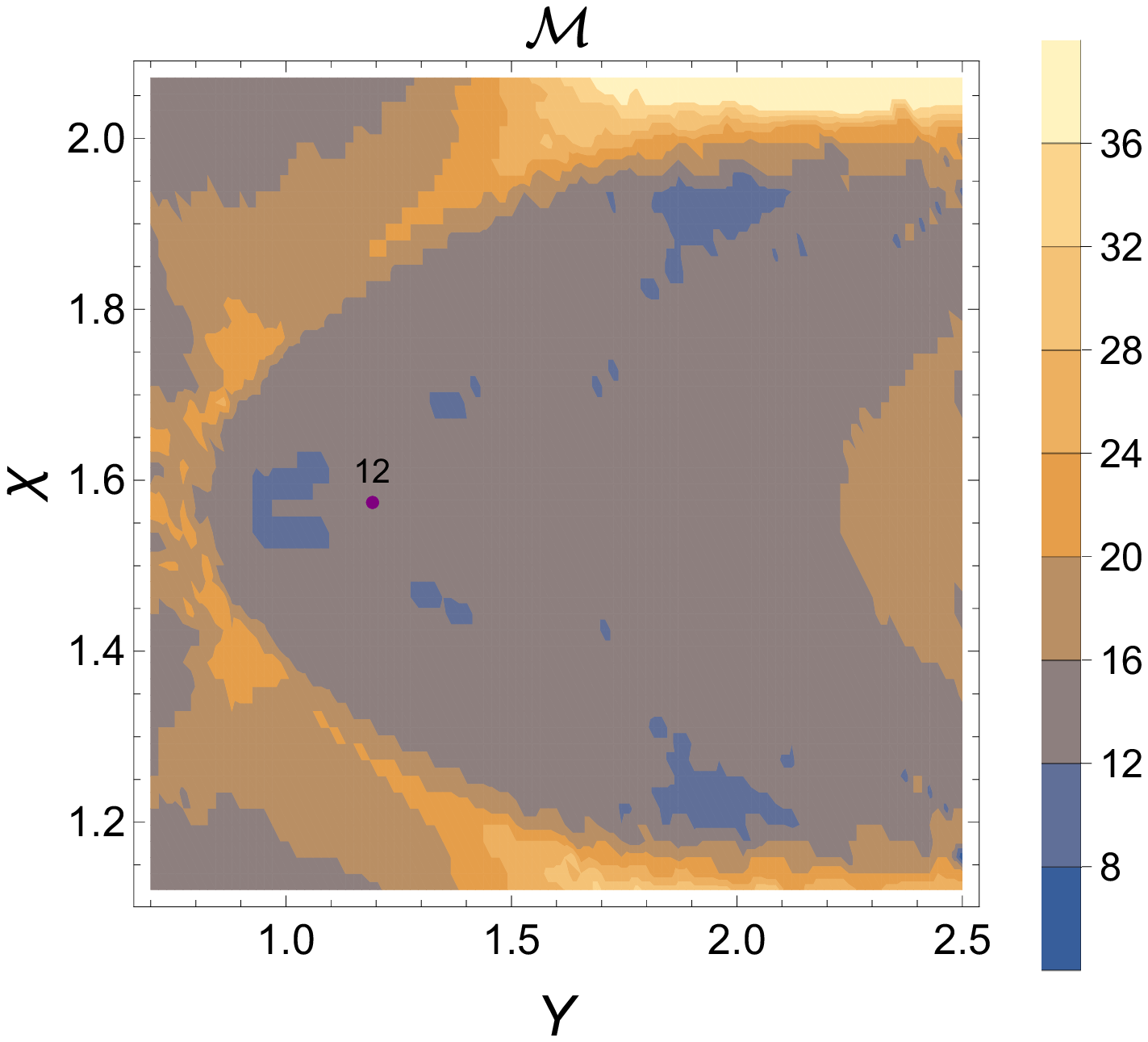}
 \caption{}
\end{subfigure}
\hfill
\begin{subfigure}[h]{0.5\linewidth}
\includegraphics[scale=0.6]{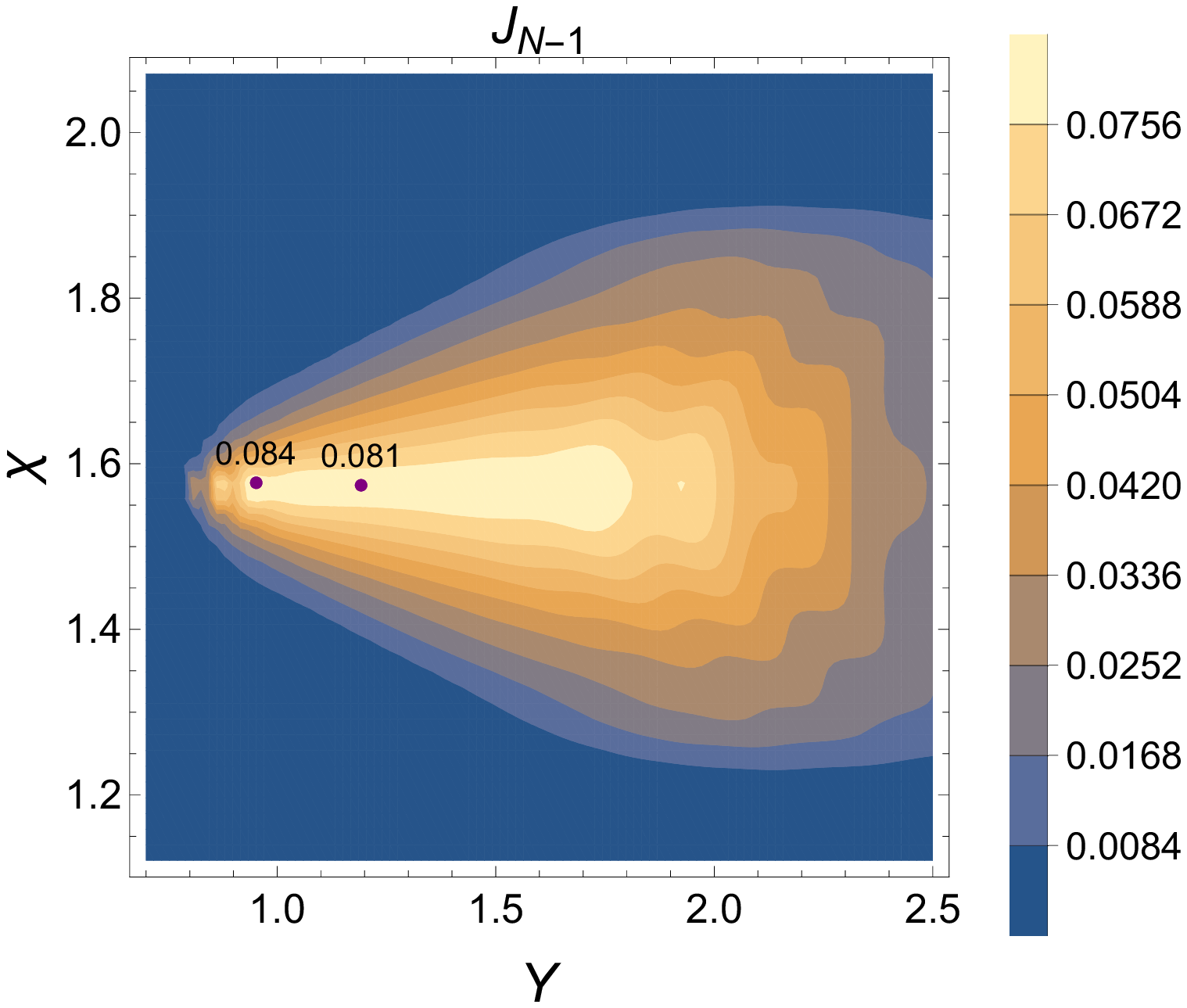}
 \caption{}
\end{subfigure}
\caption{{The zigzag chain of $N=41$ nodes, $\varepsilon=0.01$. (a) The quantity   ${\cal{M}}$ over the region (\ref{Ychi41}) of the plane $(Y,\chi)$,
the minimal value of ${\cal{M}}^{(min)}= 4$, the  point ${\cal{M}}^{(opt)}=12$  at $(Y^{(opt)},\chi^{(opt)})=(1.192, 1.574)$ is marked.  (b) The function $J_{N-1}$   over the the same region of the plane $(Y,\chi)$; this function takes its maximal value   
$J_{N-1}^{(max)}=0.084$ at $(Y^{(max)},\chi^{(max)})=(0.952, 1.577)$ and the value 
 $J_{N-1}^{(opt)}=0.081$  at $(Y^{(opt)},\chi^{(opt)})=(1.192, 1.574)$; both $J_{N-1}^{(max)}$ and $J_{N-1}^{(opt)}$ are marked { and $J_{N-1}^{(opt)}$ is close to   $J_{N-1}^{(max)}$.}}}
\label{Fig:MIN41}
\end{figure*}

\subsubsection{Even-node chain: $N=40$}
\label{Section:zigzag40}

Numerical simulations show that   $p^{(max)}$ is large ($0.1<p^{(max)}\lesssim 0.8$) inside of the region
\begin{eqnarray}\label{Ychi40}
0.65\lesssim Y\lesssim 4.2,\;\; 1.35\lesssim \chi \lesssim 2.1
\end{eqnarray}
for $T=850$, 
as shown in Fig.\ref{Fig:p40}(a).
\begin{figure*}[!]
\hspace{-1cm}
\begin{subfigure}[h]{0.5\textwidth}
\includegraphics[scale=0.6]{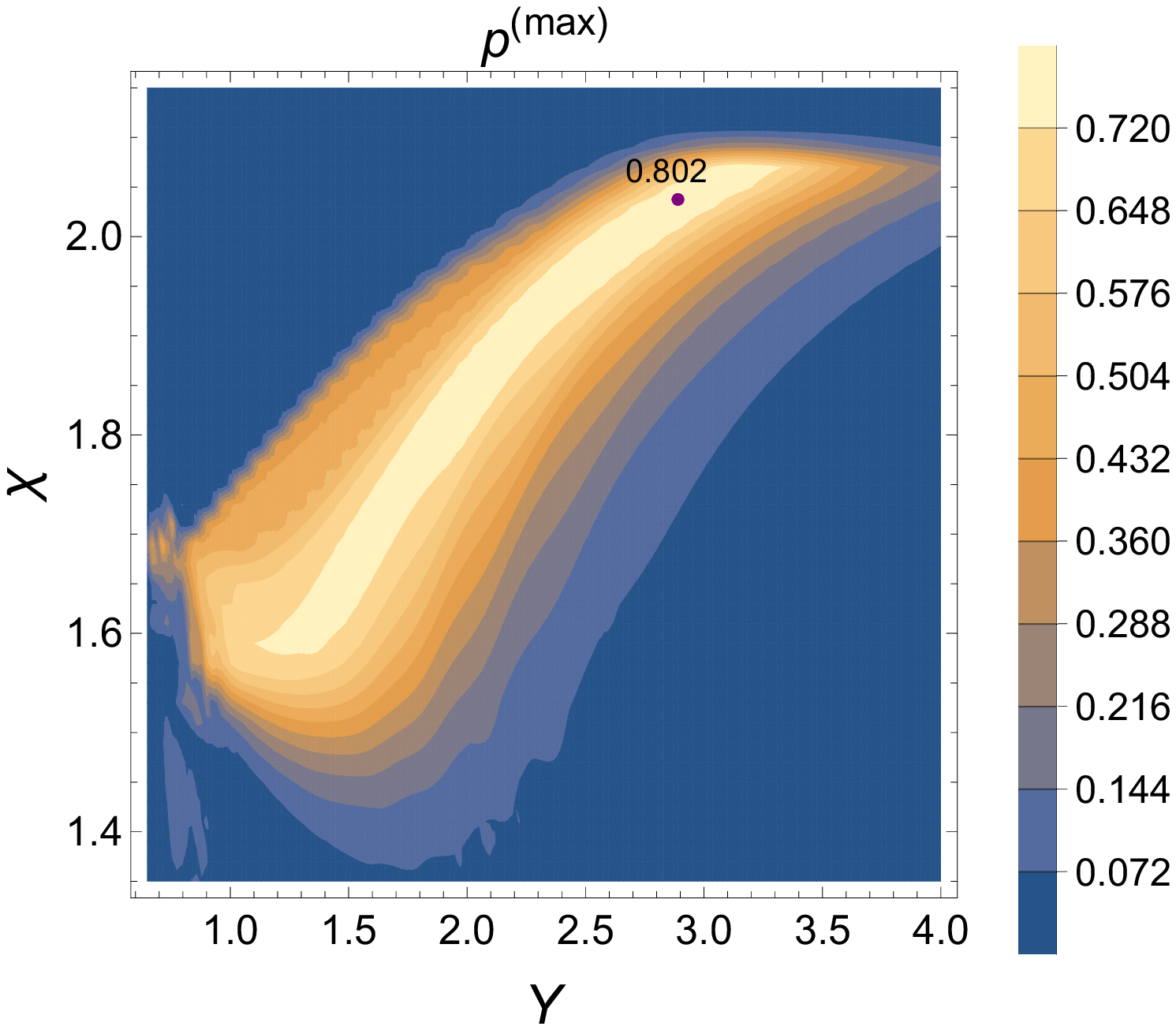}
 \caption{}
\end{subfigure}
\hfill
\begin{subfigure}[h]{0.5\linewidth}
\includegraphics[scale=0.6]{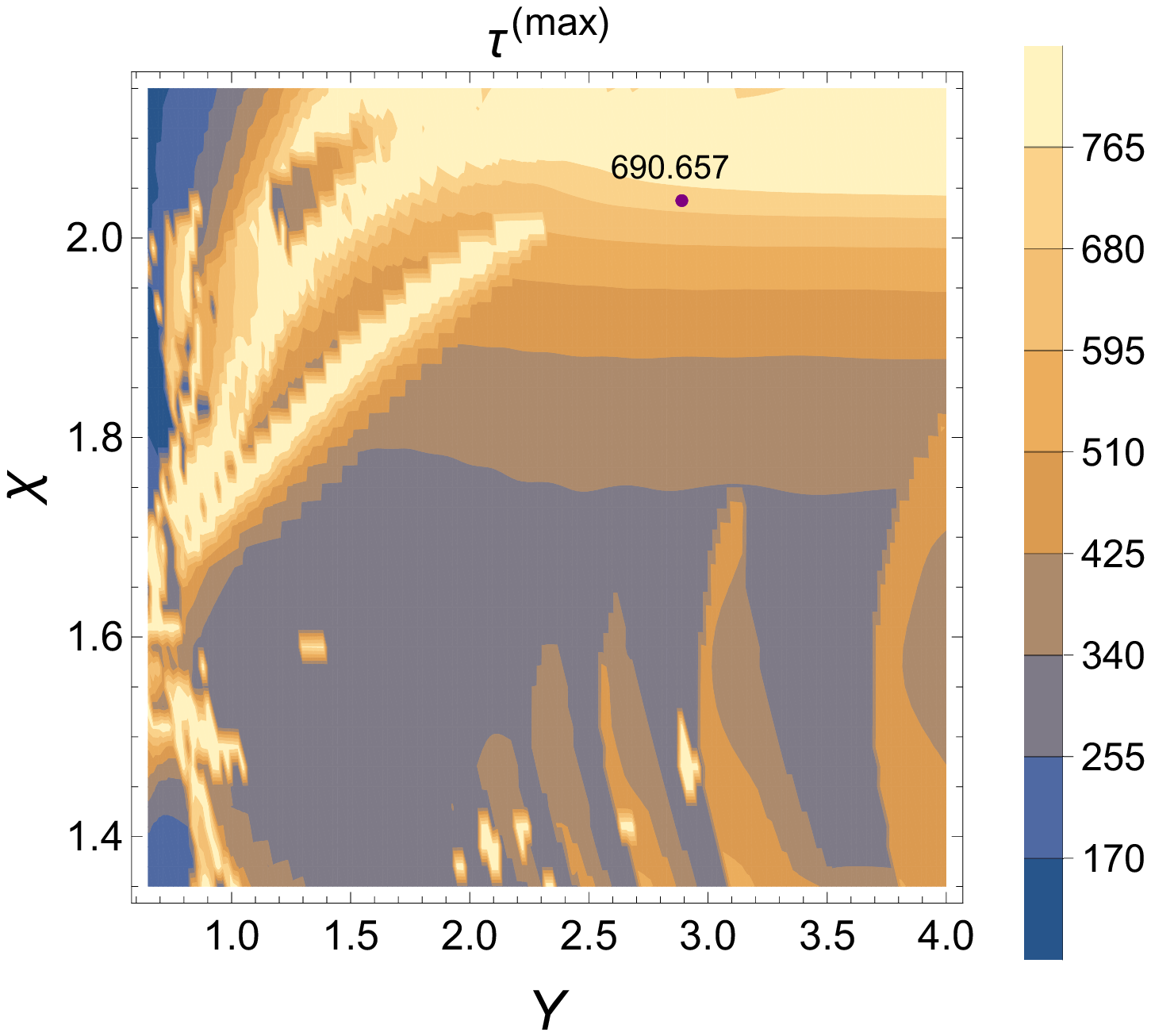}
 \caption{}
\end{subfigure}
\caption{{The zigzag chain of $N=40$ nodes. (a) The probability $p^{(max)}$  and (b) appropriate time instant $\tau^{(max)}$ as  functions 
of $Y$ and $\chi$. The probability $p^{(max)}$ reaches its maximal value $p^{(opt)}=0.802$
at $(Y^{(opt)},\chi^{(opt)})=(2.843, 2.031)$, the appropriate time instant $\tau^{(opt)}=690.657$. } }
\label{Fig:p40}
\end{figure*}
The appropriate time instant is found inside of the interval
\begin{eqnarray}\label{T40}
230 \lesssim \tau^{(max)} \lesssim 800,
\end{eqnarray}
see Fig.\ref{Fig:p40}(b).
The optimal parameters $Y^{(opt)}$, $\chi^{(opt)}$, $p^{(opt)}$ and $\tau^{(opt)}$ shown in Fig.\ref{Fig:p40} are defined  similar to the case of odd-node chain, Sec.\ref{Section:zigzag41}.

Now we find ${\cal{M}}$ according to  formulae (\ref{JM}) - (\ref{vare2}). 
The picture of ${\cal{M}}$ over the selected region (\ref{Ychi40}) on the plane $(Y,\chi)$ is shown in Fig.\ref{Fig:MIN40}a, and the 
picture of the integral $J_{N-1}$ over the same region is shown in Fig.\ref{Fig:MIN40}b. 
Thus, ${\cal{M}}$ depends on the parameters $Y$ and $\chi$. In most cases of large $p^{(max)}$ we have ${\cal{M}}\gtrsim 12$. 
We found ${\cal{M}}^{(opt)}=14$ at the optimal point 
$(Y^{(opt)}, \chi^{(opt)})=( 2.843,2.031) $. 
\begin{figure*}[!]
\hspace{-1cm}
\begin{subfigure}[h]{0.5\textwidth}
\includegraphics[scale=0.6]{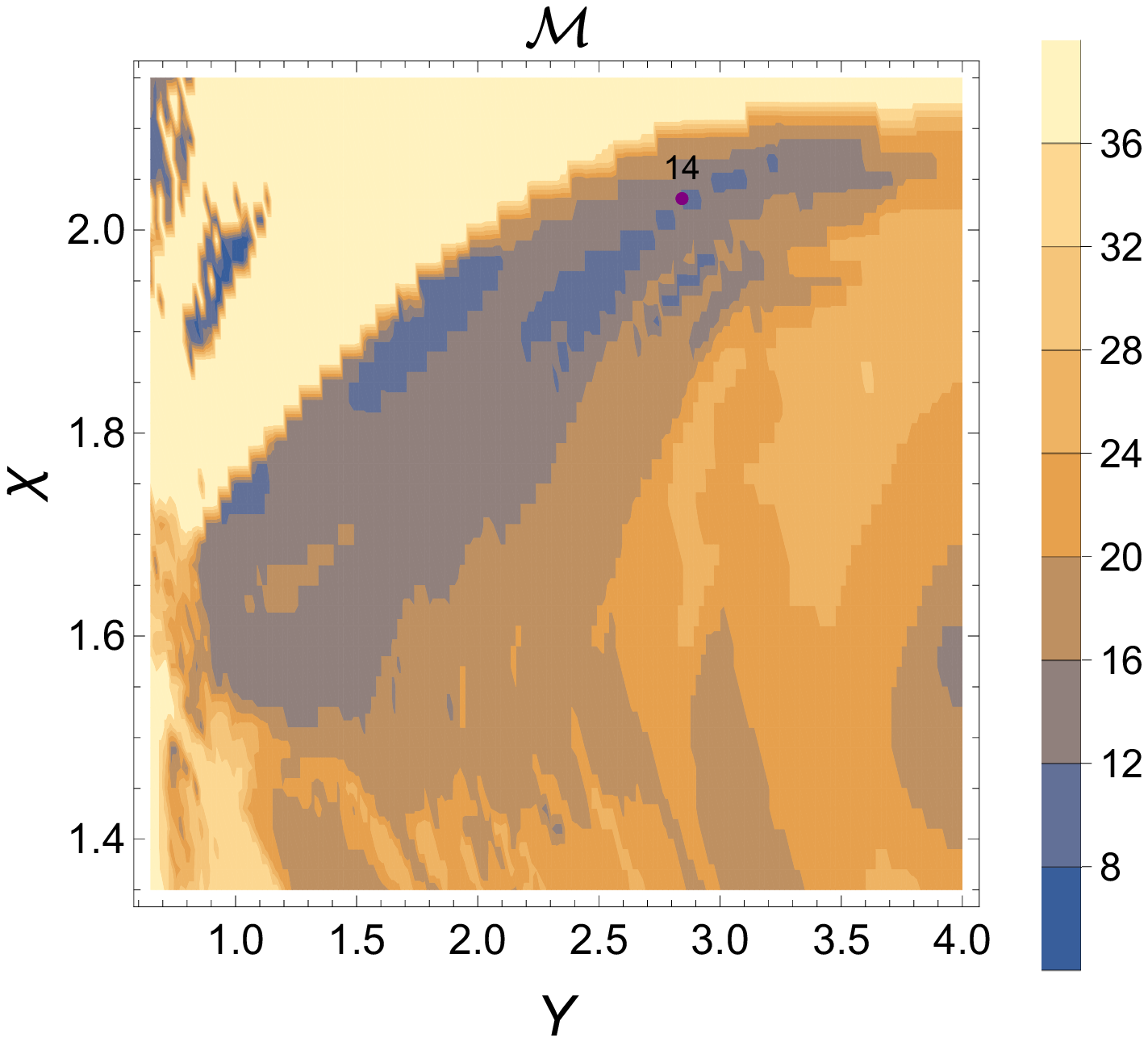}
 \caption{}
\end{subfigure}
\hfill
\begin{subfigure}[h]{0.5\linewidth}
\includegraphics[scale=0.6]{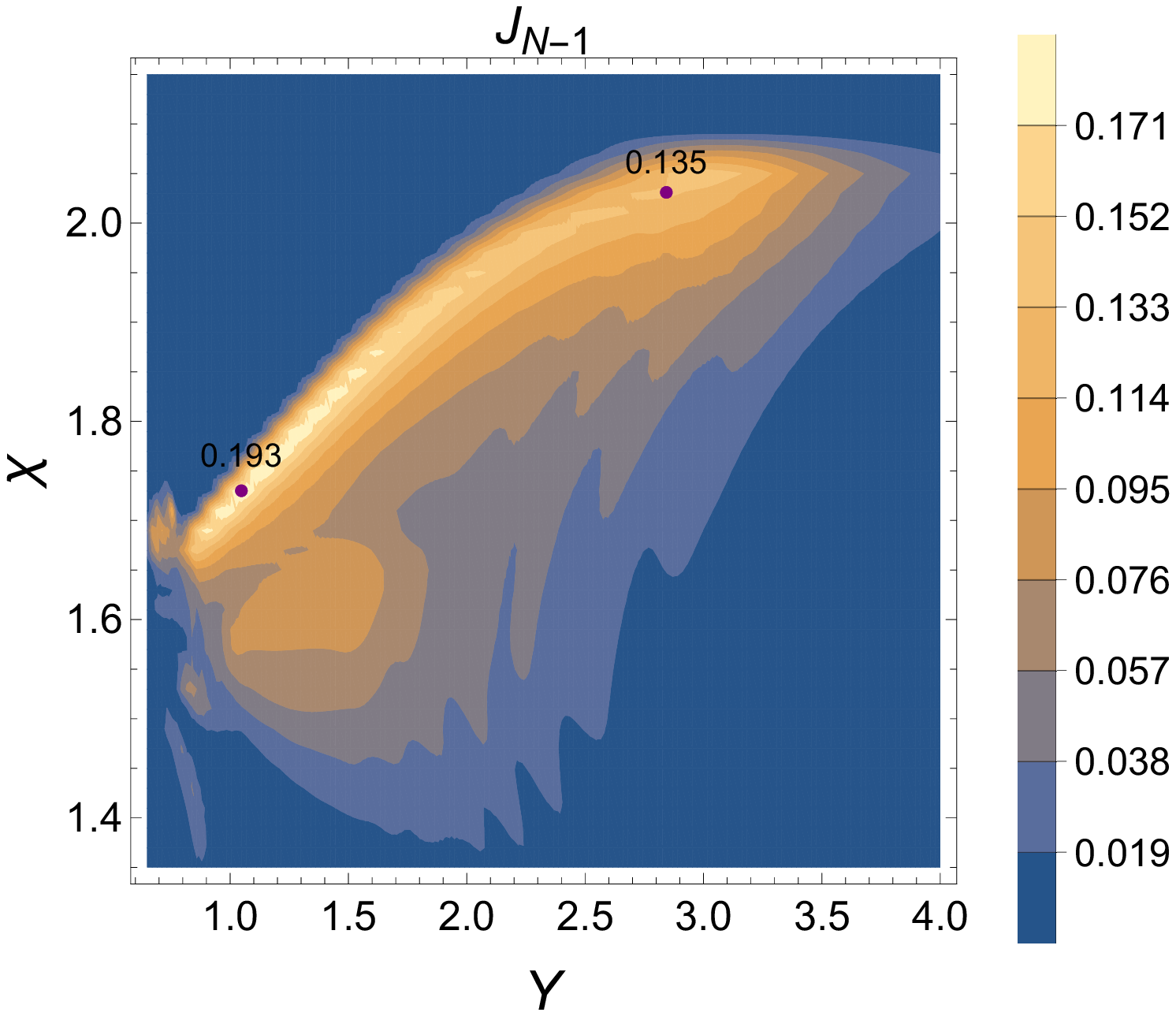}
 \caption{}
\end{subfigure}
\caption{{The zigzag chain of $N=40$ nodes, $\varepsilon=0.01$. (a) The quantity ${\cal{M}}$ over the region (\ref{Ychi40}) on the plane $(Y,\chi)$, the minimal value  ${\cal{M}}^{(min)}= 4$, the  point ${\cal{M}}^{(opt)}=14$ at $(Y^{(opt)},\chi^{(opt)})=(2.843, 2.031)$ is marked.
(b) The function $J_{N-1}$   over the same region on the plane $(Y,\chi)$, this function takes its maximal value 
$J_{N-1}^{(max)}=0.193$ at $(Y^{(max)},\chi^{(max)})=(1.05, 1.73)$ 
and the value  $J^{(opt)}_{N-1}=0.135$ at $(Y^{(opt)},\chi^{(opt)})=(2.843, 2.031)$; both $J_{N-1}^{(max)}$ and $J_{N-1}^{(opt)}$ are marked, {their values significantly differ from each other.} }}
\label{Fig:MIN40}
\end{figure*}

The graphs of the functions $p(\tau)$ for both even ($N=40$) and odd ($N=41$)  chains and  the found optimal values of the parameters $Y$ and $\chi$ are shown in Fig.\ref{Fig:pt}a. 
{We see that the state-transfer probability is  bigger for $N=40$. 
 Moreover, the profile of $p(t)$ is wider for $N=40$ which is convenient for state registration. However, the appropriate state-transfer time interval is also bigger which reduces privilege of the chain of 40 nodes over the chain of 41 nodes. }
The Graphs of the ratio $J_{M,N-1}$  for $N=40$ and $N=41$ at the 
optimal values of the parameters $Y$ and $\chi$ are shown in Fig.\ref{Fig:pt}b for $M\ge 12$. {Obviously, $M\to N-1$ with  $J_{M,N-1}\to 0$.}

{Notice that the nearest-neighbour approximation ($M=1$)  destroys the high-probability state transfer. In fact, the numerical simulation show that  $p_1\lesssim 0.05$ for all values of the parameters $Y$ and $\chi$ and both chains $N=41$ and $N=40$. We do not discuss details of such simulations.}

\begin{figure*}[!]
\hspace{-1cm}
\begin{subfigure}[h]{0.5\textwidth}
\includegraphics[scale=0.6]{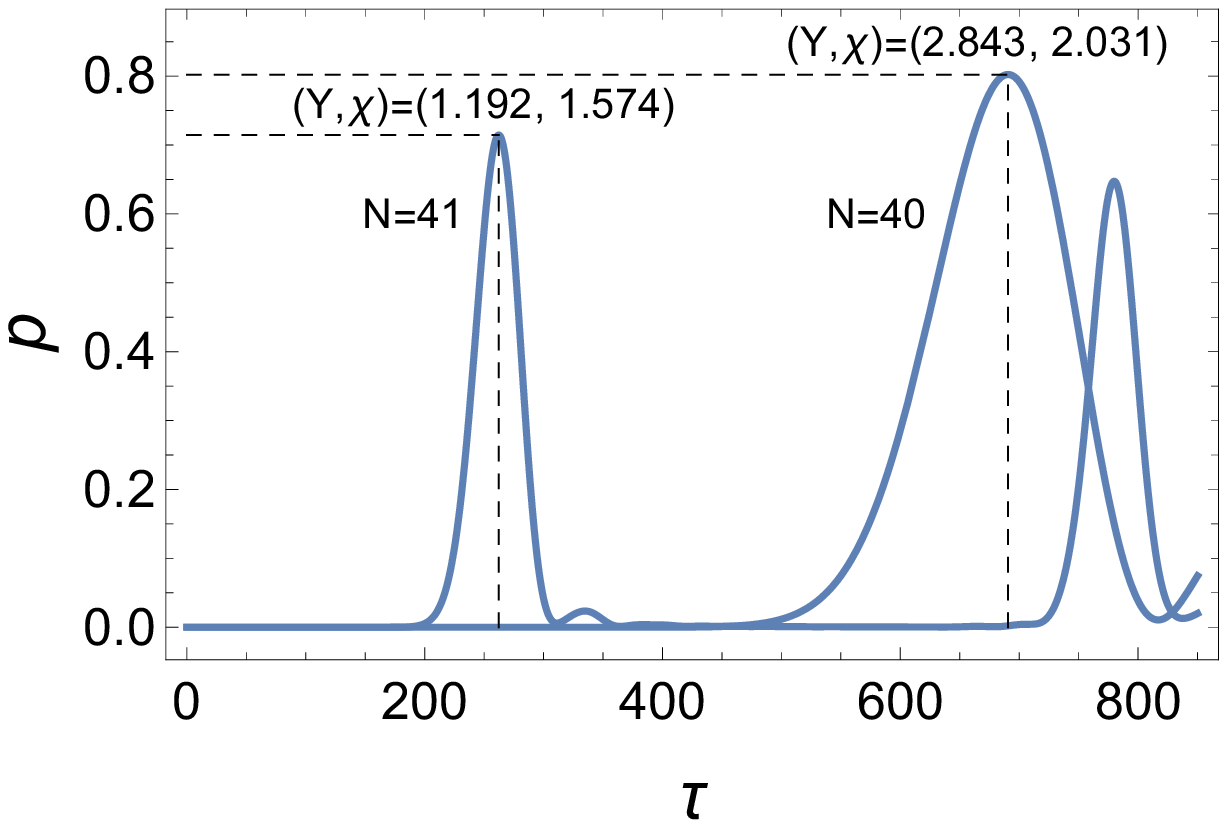}
 \caption{}
\end{subfigure}
\hfill
\begin{subfigure}[h]{0.5\textwidth}
\includegraphics[scale=0.6]{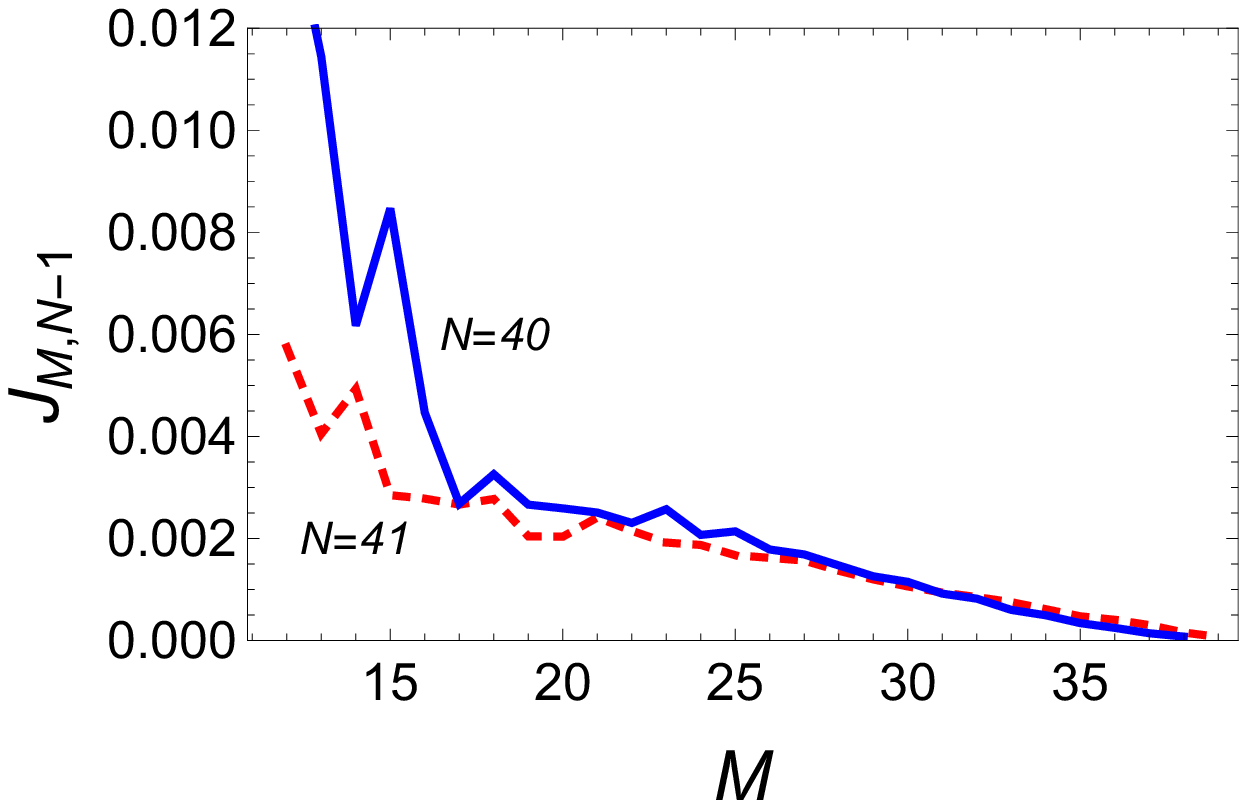}
 \caption{}
\end{subfigure}
\caption{
The zigzag spin-chains  of $N=40$ and $N=41$ nodes  at the optimal values of the parameters $Y$ and $\chi$:
$(Y^{(opt)},\chi^{(opt)})=(1.192,1.574)$ for $N=41$ and $(Y^{(opt)},\chi^{(opt)})=(2.843,2.031)$ for $N=40$.
(a) The evolution of the probability $p$ of the excited state transfer from the first to the last node;
(b) the ratio $J_{M,N-1}$ as a function of $M$.}
  \label{Fig:pt} 
\end{figure*}

\subsection{Alternating chain, Fig.\ref{Fig:Z}b}
\label{Section:alt}

 {Dealing with the alternating spin chain we   consider an even-node chain which possess the symmetry providing   high-probability state-transfer \cite{KS}}. Thus, let $N=40$ in this section. 
{According to Eqs.(\ref{altxy}), (\ref{altcoord4}), (\ref{tau2}), the geometric configuration is defined by the single parameter $\alpha$  characterizing the alternation degree of the chain. The chain orientation with respect to the external magnetic field just effects on the time-scale.} 

The maximum of the probability of the excited state  transfer $p^{(max)}$  and appropriate time instant $\tau^{(max)}$ as functions of $\alpha$ are illustrated in Fig.\ref{Fig:altpt}. Figs.\ref{Fig:altpt}c and \ref{Fig:altpt}d show the
parts of, respectively,  Figs.\ref{Fig:altpt}a and \ref{Fig:altpt}b for $1.4<\alpha < 2$. 
We see that $p^{(max)}$  approaches one in Fig.\ref{Fig:altpt}c {(almost perfect state transfer).}
\begin{figure*}[!]
\hspace{-1cm}
\begin{subfigure}[h]{0.5\textwidth}
\includegraphics[scale=0.6]{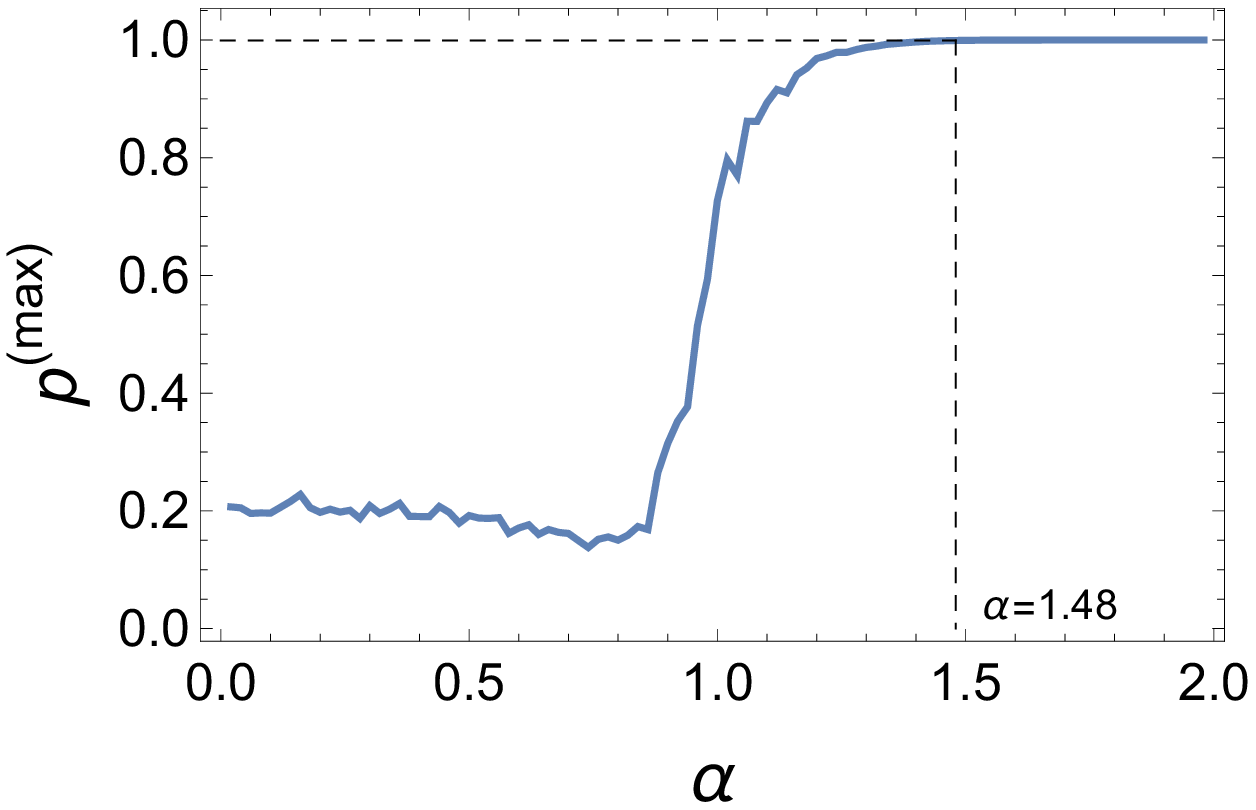}
 \caption{}
\end{subfigure}
\hfill
\begin{subfigure}[h]{0.5\linewidth}
\includegraphics[scale=0.6]{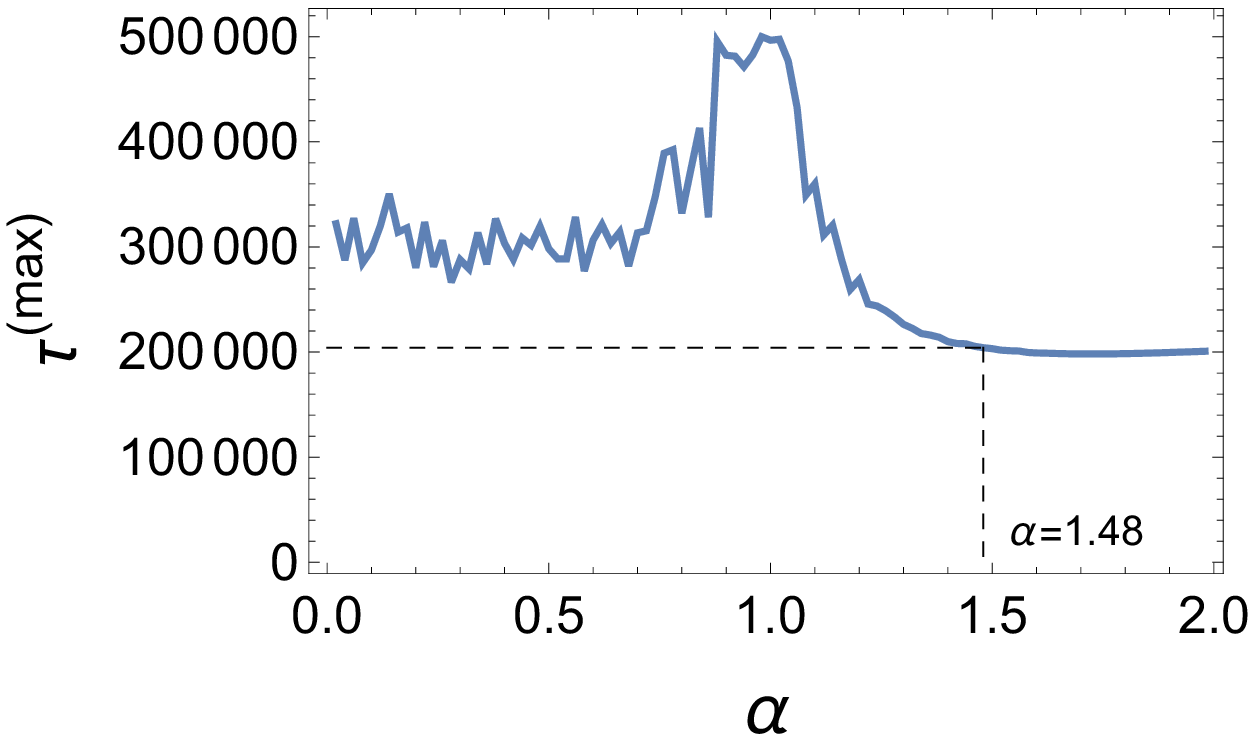}
 \caption{}
 \end{subfigure}\\\hspace{-1cm}
 \begin{subfigure}[h]{0.5\textwidth}
\includegraphics[scale=0.6]{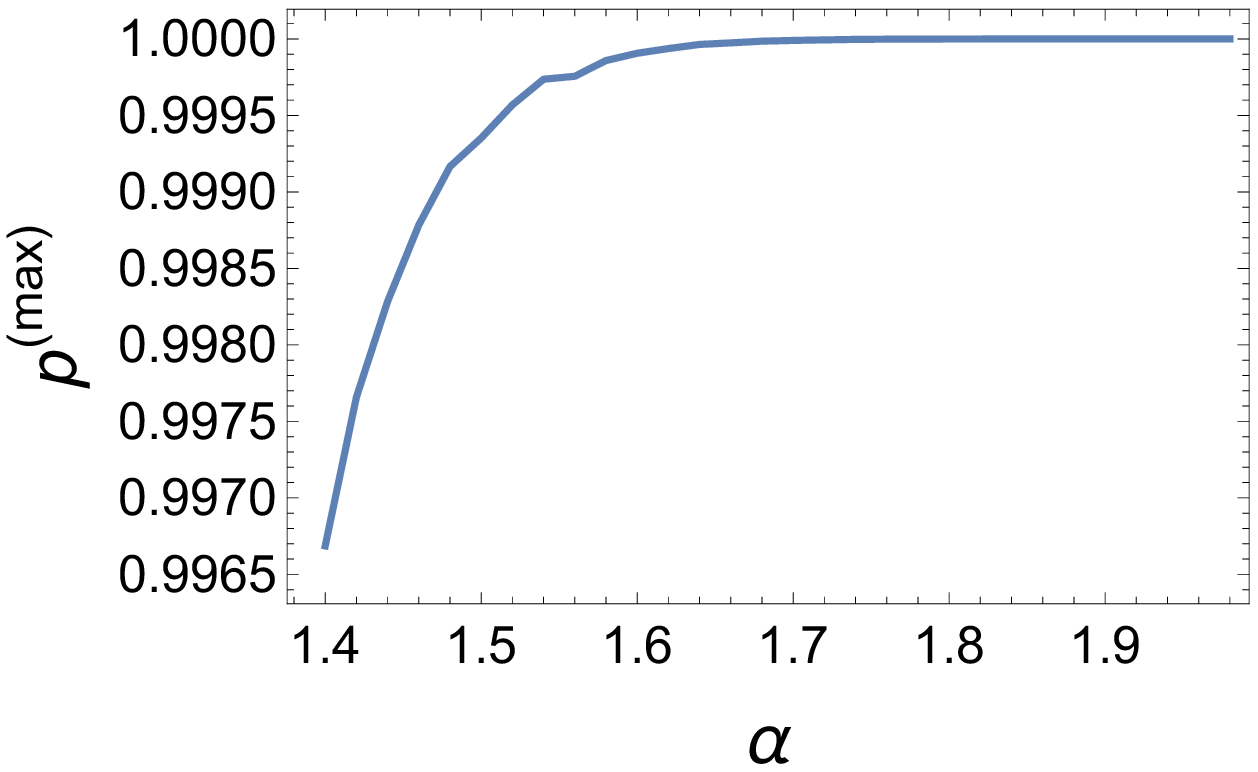}
 \caption{}
\end{subfigure}
\hfill
\begin{subfigure}[h]{0.5\linewidth}
\includegraphics[scale=0.6]{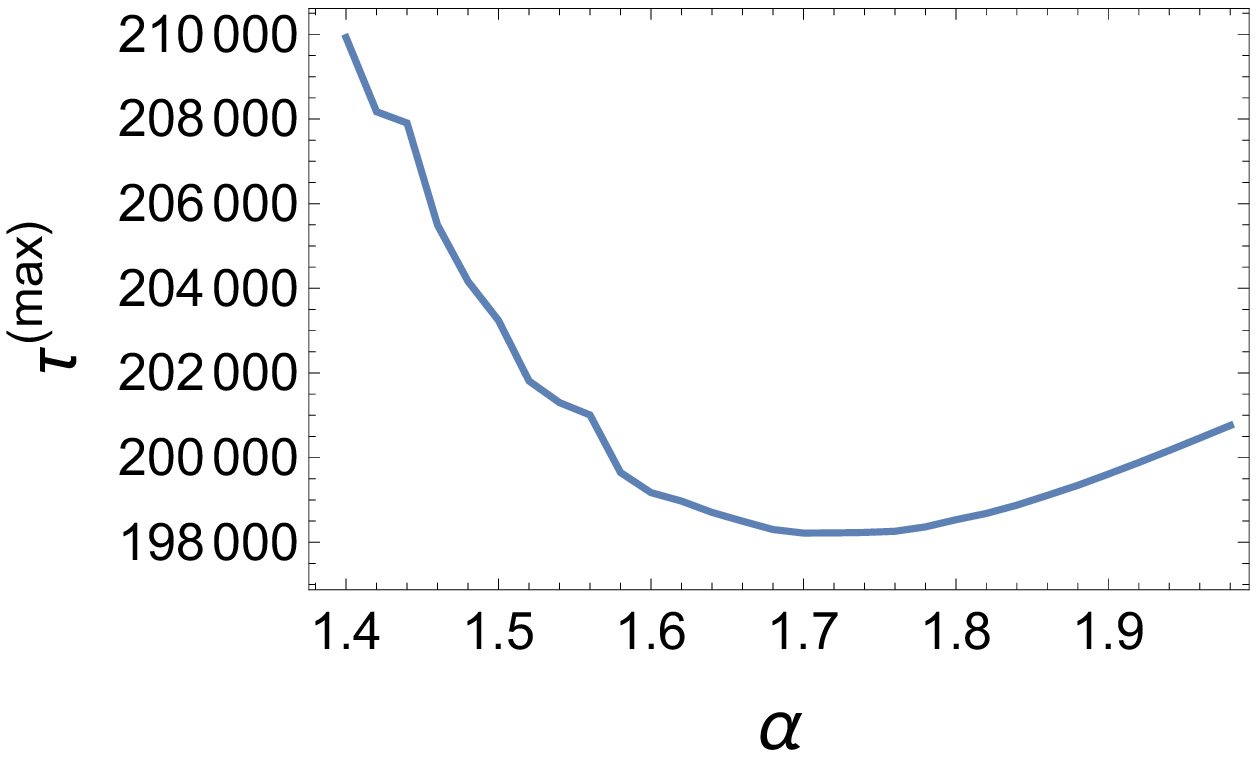}
 \caption{}
\end{subfigure}
\caption{The alternating chain with $N=40$.  (a) The probability $p^{(max)}$  and (b) appropriate time instant $\tau^{(max)}$  as  functions 
of $\alpha$. The marked point $\alpha =1.48$ corresponds to the probability  $p^{(opt)} \approx 0.999$ and  $\tau^{(opt)}=204164$. Functions $p^{(max)}(\alpha)$ and $\tau^{(max)}(\alpha)$ over the interval $1.4\le \alpha < 2$ are shown, respectively, in Figs. (c) and (d).  }
\label{Fig:altpt}
\end{figure*}
The selected value of $\alpha=1.48$ corresponds to $p^{(max)}>0.999$. The accuracy of obtained $p^{(max)}$ and $\tau^{(max)}$  in the region to the left from that point is low because of { the obstacles of calculating the global maximum of} the fast oscillating function $p(\tau)$ at $\alpha\lesssim 1.4$. On the contrary, at $\alpha \gtrsim 1.4$ the function $p(\tau)$ { looses fast oscillations and the state-transfer becomes almost perfect}. For instance, the graph of $p(\tau)$ at $\alpha=1.48$ is shown in Fig.\ref{Fig:altp}.
\begin{figure*}[!]
\epsfig{file=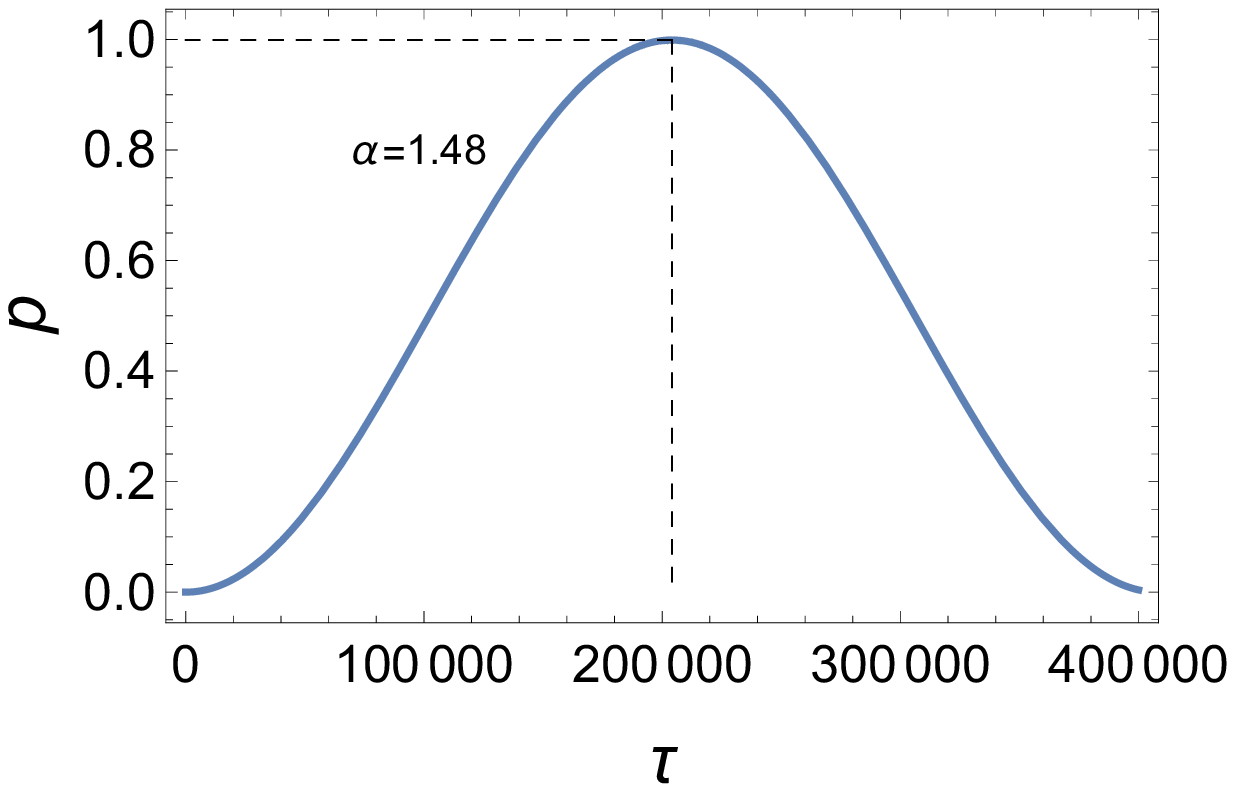,
  scale=0.7
   ,angle=0
} 
\caption{The evolution of  probability of an excited state transfer along the $N=40$-node  alternating spin chain with $\alpha=1.48$.
The maximum  $p^{(max)}\approx 0.999$ is reached  at $t^{(max)}\approx 204164$. }
\label{Fig:altp}
\end{figure*}

{
The reason of almost perfect state transfer along the alternating chain is in 
the special kind of oscillation of the state-transfer probabilities from the first to the $n$th spin, 
\begin{eqnarray}\label{P}
P_{1n}= |\langle n|V^{(1)}_{N-1}|1\rangle|^2,
\end{eqnarray}
observed in the chain, see Fig.\ref{Fig:EV}. Namely, only $P_{11}$ and $P_{1N}$ ($N=39$ in figure)  oscillate with large amplitude $\sim 1$, while all other probabilities oscillate with negligible amplitudes $\lesssim 0.0016$  (they can be hardly recognized in Fig.\ref{Fig:EV}). This can be called the Rabi-type oscillations between the one-qubit sender and receiver \cite{WLKGGB,AML}. }

\begin{figure*}[!]
\epsfig{file=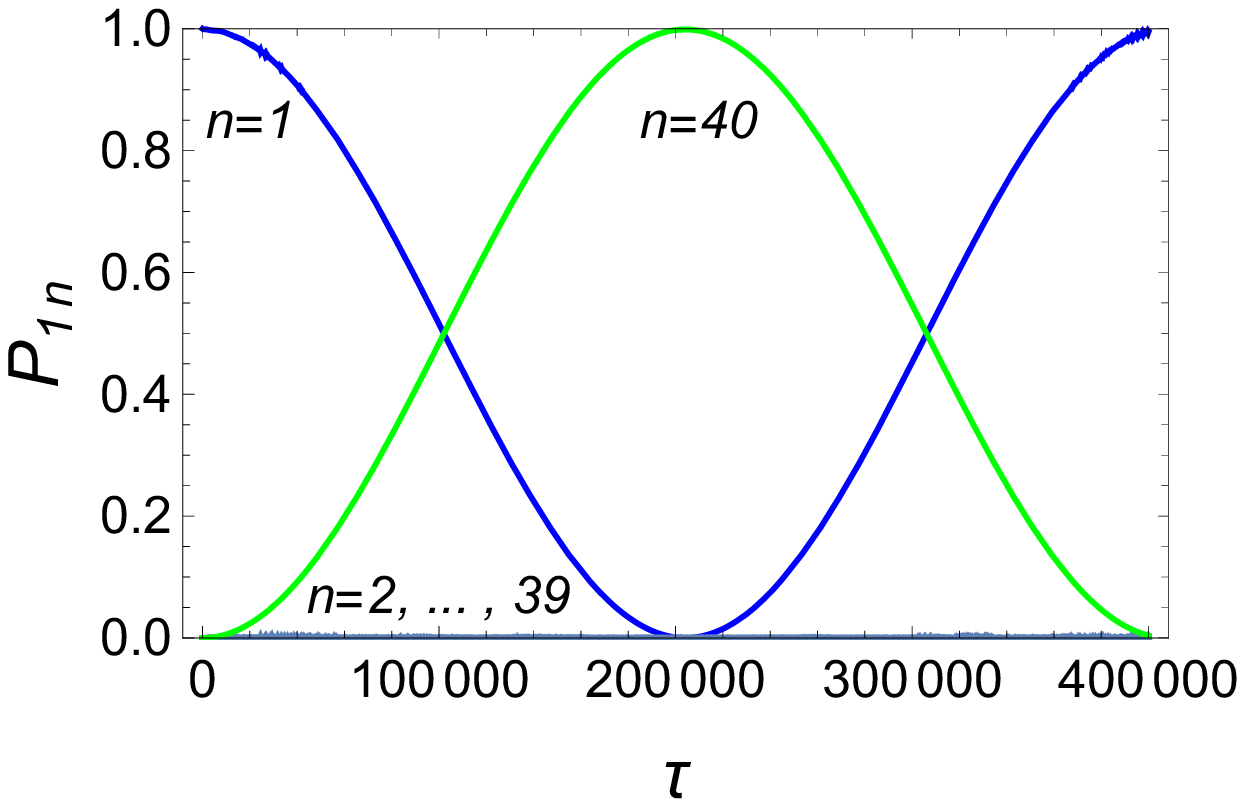,
  scale=0.7
   ,angle=0
} 
\caption{{State-transfer probabilities $P_{1n}$ defined in (\ref{P}) for $n=1$, $n=40$ (two solid lines) and $n=2,\dots,39$ (fast oscillating small-amplitude lines near the bottom).  } 
}
\label{Fig:EV}
\end{figure*}

We also study $p(\tau)$ for different $M$ in formula (\ref{XXZ}) and find out that, unlike the case of zigzag chain, the $M$ must take its maximal value $N-1$. In fact, three curves $p_M(\tau)$ for $N=40$ and $M=39$, 38 and 40 at $\alpha=1.48$ are shown in Fig.\ref{Fig:altM}. We see that reducing $M$ by 1 from $39$ to $38$ changes the period of the probability amplitude  from $\sim 5\times 10^5$ (for $p=p_{39}$)
to  $\sim 7.3\times 10^6$ (for $p_{38}$), i.e., the period becomes about 15 times longer.  Similar situation  is observed with reducing $M$ from 38 to 37, see the almost straight dotted line at the bottom of Fig.\ref{Fig:altM}. 

{ We notice that  reducing $M$ leads to narrowing in the interval of $\alpha$ corresponding to the almost perfect state transfer (i.e., $p_M\to 1$). Thus, 
this interval is large for all-node interaction and almost all-node interaction 
($1.45 \lesssim \alpha \lesssim 2$ for $M=39$, $1.32 \lesssim \alpha \lesssim 1.94$ for $M=37$). Further reducing $M$ yields $1.26\lesssim \alpha \lesssim1.38$
and 
$1.24\lesssim \alpha \lesssim1.26$ for, respectively, $M=18$ and $M=17$. For $M<17$, the almost perfect state transfer  disappears, state-transfer 
probability 
 becomes fast oscillating function with the maximum  $p^{(opt)} \lesssim 0.75$ at the time instant 
$ \tau^{(opt)} \sim 10^{10}$, which is much bigger then all state-transfer time intervals presented in Figs. \ref{Fig:p41}, \ref{Fig:p40}, \ref{Fig:altpt}. Of course, this is not the Rabi-type oscillations. Another feature  of such state transfer is that the width of the pick  is $\sim 3$, while the widths of the picks in Fig.\ref{Fig:pt} are
 $\sim 80$ for $N=41$ and $\sim  150$ for $N=40$, and this width is 
$\sim 200000$ for the 40-node alternating spin chain  in Fig.\ref{Fig:altp}. In addition, at $M=1$, the maximal probability corresponds to  $\alpha \approx 0.82<1$ (on the contrary, the maximal probability corresponds to $\alpha\gtrsim 1.5>1$ in the case of all-node interaction). Thus, the nearest-neighbor interaction  represents  a special model rather then the approximation of all-node dipole-dipole interaction.} 

\begin{figure*}[!]
\epsfig{file=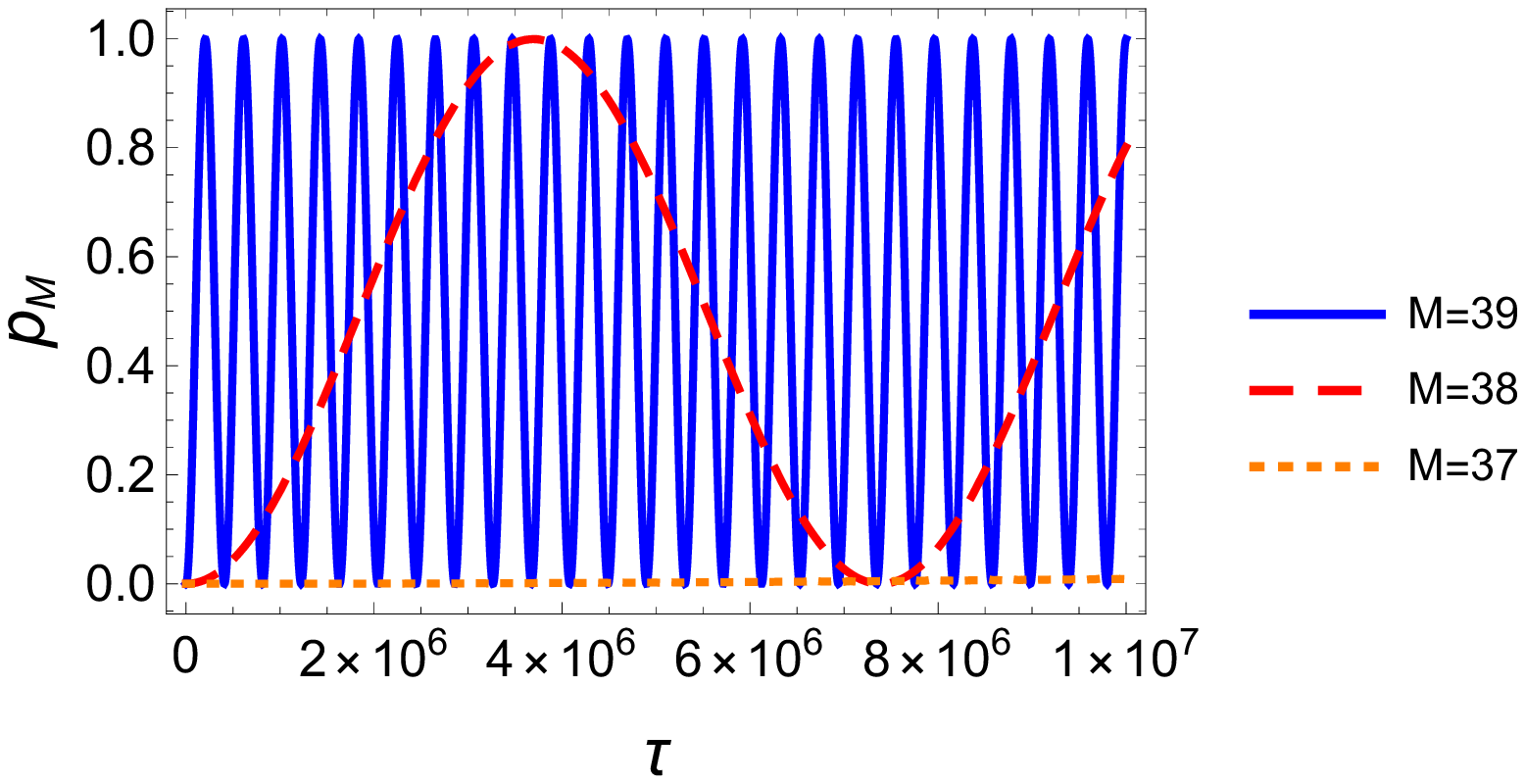,
  scale=0.7
   ,angle=0
} 
\caption{The probability $p_M(\tau)$ of the state-transfer along the  alternating chain  $N=40$ nodes 
at $\alpha=1.48$   for $M=39$, 38, 37. 
}
\label{Fig:altM}
\end{figure*}

\subsection{Comparison of results}

We collect the basic results of three cases of state transfer  considered in this paper in Table \ref{Table:1}.
\begin{table}
\begin{tabular}{|c|p{3cm}|p{3cm}|p{3cm}|}
\hline
&odd-node chain $N=41$ & even-node chain $N=40$& alternating chain\\
\hline
$p^{(opt)}$   &0.714&0.802&0.999 \\
$\tau^{(opt)}$&262.282&690.657&204164 \\
chain parameters&$Y=1.192$,\newline $\chi=1.574$& $Y=2.843$, \newline $\chi=2.031$ &$\alpha=1.48$\\
\hline
\end{tabular}
\caption{\label{Table:1}
The state transfer parameters $p^{(opt)}$ and $\tau^{(opt)}$  and the geometry parameters  $Y$, $\chi$, $\alpha$ for the zigzag and alternating chains.}
\end{table}
Thus, the probability $p^{(opt)}$ and the time instant $\tau^{(opt)}$ are both minimal for the odd-node zigzag chain and maximal for the alternating chain. The even-node zigzag chain  provides the intermediate characteristics of  the state transfer. In addition, the region on the plane $(Y,\chi)$ providing the high-probability state-transfer is rather restricted, as shown in Figs.\ref{Fig:p41}a and \ref{Fig:p40}a,  so that the successful state transfer requires accurate adjustment of the chain parameters. On the contrary, the interval for the parameter $\alpha$ providing the high-probability state-transfer along the alternating chain is large and, moreover, the state-transfer probability is almost constant over the interval $1.4 \lesssim \alpha \lesssim 2$ which makes this chain stable with respect to  perturbations. 

Comparing the odd- and even-node zigzag chains we see that, although $p^{(opt)}$ for the even-node chain is larger than that for the odd-node chain, the time instant $\tau^{(opt)}$  for the odd-node chain is more than twice shorter than that for the even-node chain. Therefore, the advantage in $p^{(opt)}$ is compensated by the disadvantage in $\tau^{(opt)}$. 

Thus, the zigzag chain can serve for the fast state transfer with low probability, which means essential deformation of the transferred state. Therefore, the methods for restoring the parameters of the transferred state are of great interest. 

 We notice  that, although all three graphs $p(\tau)$ are bell shaped,   the shape of $p(\tau)$  for the alternating chain is wider, compare  Fig.\ref{Fig:pt}a with Fig.\ref{Fig:altp}. Therefore, the alternating chain is less sensitive to the particular  time instant for state registration at the receiver side.

Finally, we emphasize that $p(\tau)$ for zigzag and alternating chain exhibit different dependence on the number of interacting neighbors (the parameter $M$ in Eq.(\ref{XXZ})). To obtain the good approximation to the case of all-node interactions in zigzag chain with 41 and 40 nodes we can take, respectively, $M=12$ and $M=14$  according to Sec.\ref{Section:zigzag41}, \ref{Section:zigzag40}. On the contrary, obtaining the correct probability amplitude for alternating chain requires including the all-node interaction ($M=N-1$), as shown in Sec.\ref{Section:alt}. 
 
Thus, all three considered cases have certain advantages and disadvantages and their application depends on the particular requirements to the state transfer.

\section{Conclusions}
\label{Section:conclusions}
{ In this paper we formulate the problem of $M$-neighbor approximation to the dipole-dipole interaction and investigate it for the spin-1/2 system  governed by the $XXZ$ Hamiltonian.  For this purpose we consider  the one-qubit  excited pure  state transfer along the zigzag chain with either odd or even number of nodes taking into account all-node interaction. Then we consider the possibility to correctly describe the above spin-dynamics using $M$-neighbor approximation and find appropriate parameter $M$.  The obtained state-transfer characteristics are compared with those for the alternating chain. 

Resuming our study, we show that the $M$-neighbor approximation for zigzag ($N=40,\;41$) and alternating ($N=40$) chains works, respectively,  at $M=12$, $M=14$ and $M=39$. Thus, the approximation to nearest-node interaction ($M=1$) is not applicable to the chains with dipole-dipole interaction (at least, governed by $XXZ$ Hamiltonian)  and, moreover, the evolution along the  alternating chain requires all node interaction. Thus we demonstrate the restricted applicability of $M$-neighbor  approximation to the dipole-dipole interaction. This means that, generically,  all-node interaction must be relied on  unless the particular value of $M$ is revealed for the   process of our interest in advance. 
}

As another result, we show that the high-probability state-transfer via a zigzag chain  can be observed at the time instant $\sim 10^3$ times less then the time instant of state transfer along the alternating spin chain with $\alpha \sim 1.5$. However, the probability of state transfer along the alternating chain is higher and approaches unit. 
We demonstrate that the geometry and orientation of the spin-chain are a privileged characteristics of the communication line and strongly effect both state-transfer probability (and, consequently, fidelity)   and state-transfer time. The zigzag chain is attractive due to the shorter state-transfer  time interval  in comparison with the alternating chain. Since the probability $p^{(opt)}$ of such transfer  is far from unit, the method of state restoring might be helpful to raise the effectiveness of the state-transfer protocol. 

{ We shall also notice that the almost perfect state transfer assotiated with the Rabi-type oscillations can be reached in even-node zigzag chain over about the same time interval as for the 
alternating chain (for instance, $\tau\sim 4.5\times 10^5$ for $N=40$ at $y=0.3$, $\chi = 2.513$). We do not explore this case as far as zigzag chain has no preferences over the alternating chain regarding the long-time state transfer.
}

We acknowledge funding from the Ministry of Science and Higher Education of the Russian Federation (Grant No. 075-15-2020-779).

\end{document}